\documentclass[11pt,a4paper]{article}

\usepackage{xcolor}
\usepackage[left= 2cm, right= 2cm]{geometry}
\usepackage{amsmath,latexsym,amssymb,slashed}
\usepackage{graphicx}
\usepackage{hhline}
\usepackage[titletoc]{appendix}
\usepackage{amssymb}
\usepackage{tabularx}				
\usepackage[vcentermath]{youngtab}	
\usepackage{float}
\usepackage[latin1]{inputenc}
\usepackage[T1]{fontenc}
\usepackage{slashed}
\usepackage{hyperref}
\usepackage{comment}
\usepackage{ dsfont }

\usepackage[vcentermath]{youngtab}

\newdimen\squaresize \squaresize=12pt
\newdimen\thickness \thickness=0.7pt

\def\square#1{\hbox{\vrule width \thickness
   \vbox to \squaresize{\hrule height \thickness\vss
      \hbox to \squaresize{\hss#1\hss}
   \vss\hrule height\thickness}
\unskip\vrule width \thickness} \kern-\thickness}

\def\cut#1{\hbox{\vrule width-1 \thickness
   \vbox to \squaresize{\hrule height-1 \thickness\vss
      \hbox to \squaresize{\hss#1\hss}
   \vss\hrule height-1\thickness}
\unskip\vrule width +4 \thickness} \kern-\thickness}

\def\vsquare#1{\vbox{\square{$#1$}}\kern-\thickness}

\def\young#1{
\vbox{\smallskip\offinterlineskip \halign{&\vsquare{##}\cr #1}}}

\newcommand{\tinyyoung}[1]{
\squaresize=7pt \thickness=0.4pt \mbox{\scriptsize\young{#1}}
\squaresize=15pt \thickness=0.7pt}

\numberwithin{equation}{section}

\usepackage{cite}

\newcommand{\rd}{{\mathrm d}}

\usepackage{makecell}
\usepackage{colortbl}

 \usepackage{setspace}

\usepackage{hyperref}
 \hypersetup{
    colorlinks,%
    citecolor=magenta,%
    filecolor=blue,%
    linkcolor=magenta,%
    urlcolor=blue
}

\definecolor{MyDarkBlue}{rgb}{0,0,0.7}

\definecolor{MyDarkRed}{rgb}{0.7,0,0}

\definecolor{MyDarkGreen}{rgb}{0,0.7,0}

\newcommand{\eq}[1]{\begin{equation}#1\end{equation}}
\newcommand{\spl}[1]{\begin{split}#1\end{split}}

\def\d{\text{d}}

\title{Draft Article}
\author{Alfred Bovon}

\begin{document}

\begin{titlepage}
\vfill
\begin{flushright}
\end{flushright}

\vfill
\begin{center}
	{\LARGE \bf Holographic deformations of matrix models
	}\\[1cm]
	
{\bf Alfred Bovon\,$^{a}{\!}$
		\footnote{\tt a.bovon@ip2i.in2p3.fr}, Henning Samtleben\,${}^{b,c}{\!}$
		\footnote{\tt henning.samtleben@ens-lyon.fr}, Dimitrios Tsimpis\,${}^{a}{\!}$
		\footnote{\tt tsimpis@ipnl.in2p3.fr}
		}
\vskip .8cm
	
	{\it ${}^a$ Institut de Physique des Deux Infinis de Lyon (IP2I),\\
Universit\'e de Lyon, UCBL, CNRS/IN2P3, UMR 5822,\\
4 rue Enrico Fermi, 69622 Villeurbanne Cedex, France}\\[2ex]
	{\it  $^{b}$ ENS de Lyon, CNRS, LPENSL, UMR5672, 69342, Lyon cedex 07, France}\\[2ex] 
		{\it  $^{c}$ Institut Universitaire de France (IUF), Paris, France}
	
\end{center}
\vfill

\begin{center}
	\textbf{Abstract}
	
\end{center}
\begin{quote}
We study maximal supergravity in two dimensions, obtained from reduction of IIA supergravity on an $S^8$ sphere. The theory captures the low-lying
fluctuations around the non-conformal D0-brane near-horizon geometry, dual to operators in the BFSS matrix model. Upon exciting some of the supergravity scalars, we construct half-supersymmetric domain wall solutions preserving $SO(p)\times SO(9-p)$ subgroups of the original $SO(9)$ symmetry. We determine their uplift to ten dimensions and the corresponding distributions of D0-branes. Finally, we compute the fluctuations  around these domain wall backgrounds, corresponding to holographic two-point  correlation functions in the Coulomb branch of the matrix model.

\end{quote}
\vfill
\setcounter{footnote}{0}

\end{titlepage}

\begin{spacing}{0.8}
{
  \hypersetup{linkcolor=black}
  \tableofcontents
}
\end{spacing}

\newpage

\section{Introduction}

The low-energy world-volume theory of $N$  D$0$-branes in IIA string theory is an $\mathcal{N}=16$ supersymmetric  Yang-Mills theory in  0+1 dimensions, with $U(N)$ gauge group  (or $SU(N)$, upon decoupling the $U(1)$ center-of-mass sector) and $SO(9)$ global symmetry \cite{Witten:1995im, Townsend:1995af}. 
This theory, also known as ``matrix theory'' or ``(BFSS) matrix model'', 
can also be obtained as a  matrix regularization of the M-theory supermembrane \cite{Hoppe:1986aj, deWit:1988wri}. 
It was proposed  in \cite{Banks:1996vh} as a non-perturbative formulation of M-theory with asymptotically flat boundary conditions, in an appropriately defined  $N\rightarrow\infty$ limit. 

 Alternatively, one can take  the standard decoupling limit \cite{Maldacena:1997re}, whereby the matrix model admits an equivalent (dual) description as the near-horizon limit of  $N$  coincident D$0$-branes, captured by the near-horizon geometry  of a charged black hole 
in ten-dimensional (10d) IIA supergravity \cite{Horowitz:1991cd}. 
Thereby one obtains an example of non-conformal gauge-gravity duality 
\cite{Itzhaki:1998dd}, for which holographic renormalization proceeds similarly to the conformal case \cite{Wiseman:2008qa,Kanitscheider:2008kd}. 
This is the setup that we will be the main  focus of  the present paper.

Unlike the conformal case, both the curvature and the  dilaton depend  on the radial coordinate $r$ of spacetime, reflecting the running of the 
effective coupling constant in the dual gauge theory. 
The 10d supergravity description can only be trusted for intermediate values of $r$, 
in a regime where   $N$ is large and the gauge-theory side of the duality is strongly coupled.\footnote{Tests of the gauge/gravity correspondence at weak gauge coupling are rare, see e.g. \cite{Sekino:2019krf}.}  This has motivated developing fully non-perturbative, numerical methods in order to study the matrix model  
\cite{Catterall:2007fp,Anagnostopoulos:2007fw, Catterall:2008yz,Hanada:2008gy, Hanada:2011fq, Hanada:2013rga, Kadoh:2015mka, Filev:2015hia, Berkowitz:2016jlq, Rinaldi:2017mjl}
and related deformations  \cite{Costa:2014wya,Pateloudis:2022ijr}  in the strong coupling regime.

The matrix model includes nine $N\times N$  matrices $X^i$,  $i=1,\dots, 9$,  transforming in the adjoint representation of $SU(N)$  and in the vector representation  of the $SO(9)$ R-symmetry group. 
Non-conformal holography for D0 branes has been primarily performed in the $SO(9)$-invariant vacuum of the theory, where the vacuum expectation values (VEV) of all the $X^i$ are vanishing. 
On the supergravity side this corresponds to $N$ coincident D0 branes, giving rise  to a 10d  geometry which is conformally AdS$_2\times S^8$ 
in the near-horizon limit. 
Upon uplift to 11d supergravity and an infinite Lorentz boost along the M-theory circle, the geometry becomes that of a pp-wave background, see e.g.~\cite{Taylor:2001vb, Polchinski:1999br} for reviews. 
A dictionary between 
Kaluza-Klein (KK) compactification of IIA supergravity on this background and its 11d uplift, and matrix model operators has been established in \cite{Sekino:1999av, Sekino:2000mg}.

A finite subset of these KK modes is captured by the 2d maximally-supersymmetric $SO(9)$ gauged supergravity constructed in \cite{Ortiz:2012ib}.  Subsequently,  it was shown in \cite{Anabalon:2013zka} that a $U(1)^4$ truncation thereof is a consistent truncation of IIA, and its uplift to ten dimensions was explicitly constructed. In particular, the conformal  AdS$_2\times S^8$  near-horizon geometry of $N$  D0 branes was recovered as the uplift to ten dimensions of an $\mathcal{N}=16$ supersymmetric  domain-wall solution of the 2d theory. More recently, based on exceptional field theory techniques \cite{Bossard:2018utw, Bossard:2021jix}, the full 2d maximally-supersymmetric $SO(9)$ gauged supergravity was shown to be a consistent truncation of 11d supergravity \cite{Bossard:2022wvi, Bossard:2023jid, Bossard:2023wgg}.

In a previous paper \cite{Ortiz:2014aja} we used a subsector of the 2d maximally-supersymmetric gauged supergravity to perform holographic calculations in an $SO(3)\times SO(6)$ invariant vacuum of the theory. 
Recall that separating some of the D0 branes in the transverse dimensions  corresponds in the matrix model to  breaking the  $SO(9)$ R-symmetry by giving non-trivial VEV's to some of the  $X^i$, 
subject to the flatness condition $[X^i,X^j]=0$.~One thus obtains a moduli space of vacua, referred to as the ``Coulomb branch'' of the 
gauge theory. Whenever the $X^i$ can  be simultaneously diagonalized, it is  meaningful to talk about the positions of the D0 branes in the transverse space, given by the corresponding  eigenvalues. Within the supergravity approximation in the large-$N$ limit, 
one expects  in this case to have a description in terms of continuous distributions of branes, and the space of all such configurations  to correspond to the  Coulomb branch of the gauge theory \cite{Kraus:1998hv, Freedman:1999gk}.

However, the situation described in the previous paragraph, following the  standard conformal holography intuition,  
runs into several subtleties in the case of the BFSS model.~Firstly, it should be noted that for finite $N$ all energy eigenstates except the ground state are scattering states.~This is the  holographic dual of the statement that the black hole is unstable to decay by emission of D0-branes \cite{Lin:2013jra}.~On the other hand, the decay is exponentially suppressed for large $N$, and we  may argue that  this subtlety can be ignored \cite{Biggs:2023sqw}.~Secondly, evidence from Monte Carlo simulations suggests that the commutator $[X^i,X^j]$ can be large in the matrix model,\footnote{An upper bound for the quantity $0\leq\langle|\text{Tr}[X,Y]|^2\rangle/\langle\text{Tr}X^2\rangle\langle\text{Tr}Y^2\rangle$ was provided in \cite{Lin:2023owt}.} rendering   the notion of a Coulomb branch problematic. 
We will use the term ``Coulomb branch'' to refer to the duals of the supergravity solutions discussed in the present paper, because, as we will 
argue, they should be identified with VEV deformations  of the standard SO(9) vacuum of the matrix model at zero temperature.~This interpretation will be further supported by the fact that, as we will see,  
our supergravity solutions uplift to multi-center brane distributions. 

Generic points in the Coulomb branch should correspond to supergravity configurations that asymptote the conformal AdS$_2\times S^8$ geometry far from the D0 branes, and  involve  an   infinite number of non-trivial KK mode profiles. 
In  \cite{Ortiz:2014aja} we  considered a distinguished $SO(3)\times SO(6)$ invariant point in the Coulomb branch that corresponds to exciting only  certain scalar modes captured by the 2d maximal supergravity. 
In the present paper we generalize this work to $SO(p)\times SO(9-p)$ invariant vacua of  2d maximal supergravity,  for all $p=1,\dots,8$, by  explicitly constructing  the corresponding D0 brane distributions 
and their  uplifts   to eleven-dimensional pp-waves. 

Furthermore, we set up the holographic renormalization scheme to compute two-point correlation functions, in the Coulomb branch of the matrix model, of the operators dual to the scalar fields in the different representations of the   $SO(p)\times SO(9-p)$ decompositions
of $SO(9)$. In the UV-limit, all correlators are consistent with those computed in the $SO(9)$ vacuum of the matrix model. Our results can in principle be used as highly nontrivial tests of the holographic duality, by comparing to independent  numerical calculations in a regime where the matrix quantum mechanics is strongly coupled.

\section{\texorpdfstring{$SO(p)$}{SO(p)}~\texttimes \texorpdfstring{$~SO(q)$}{SO(q)} domain wall solutions}

\subsection{\texorpdfstring{$SO(9)$}{SO(9)} supergravity and its domain wall}

The two-dimensional maximally supersymmetric SO(9) supergravity was constructed in \cite{Ortiz:2012ib}. As shown in the series of papers~\cite{Bossard:2022wvi,Bossard:2023jid,Bossard:2023wgg}, the theory describes the consistent truncation of IIA supergravity on the sphere $S^8$. As such, it provides a convenient tool to construct solutions of IIA supergravity and moreover to  compute the fluctuations of the supergravity multiplet around these solutions. In two dimensions, the theory carries a metric $g_{\mu\nu}$ together with a dilaton $\rho$ and 128 scalar fields, transforming as \textbf{44} $\oplus$ \textbf{84} under SO(9). We describe the scalars in the \textbf{44} by a symmetric ${SL}(9)$ matrix $T_{IJ}$, $I, J=1, \dots, 9$, parametrising the coset space $SL(9)/SO(9)_{K}$. The scalar fields transforming in the \textbf{84} under $SO(9)$, are denoted $a^{IJK}$. 
In addition, the theory carries the $SO(9)$ vector fields $A_\mu{}^{IJ}$ with non-abelian field strengths $F_{\mu\nu}^{IJ}$, as well as the auxiliary fields $Y^{IJ}=-Y^{JI}$ in the adjoint representation of $SO(9)$.

The SO(9) supergravity Lagrangian reads
\begin{align}\label{eq:gaugedLag}
\mathcal{L}_{\rm sugra} =\ & 
\sqrt{-g}\left( -\rho R - \frac14\,\rho  g^{\mu\nu}  D_\mu T^{IJ} D_\nu T^{-1}_{IJ} 
+\frac{1}{12}\rho^{1/3}g^{\mu\nu} D_\mu a^{IJK}  D_\nu a^{LMN} T^{-1}_{IL}T^{-1}_{JM}T^{-1}_{KN} \right)  \nonumber\\&
+\frac12\varepsilon^{\mu\nu} F_{\mu\nu}^{IJ}\, Y_{IJ}  
+{\cal L}_{\rm WZ} 
- \sqrt{-g}\,V_{\text{sugra}}\,,
\end{align}
in spacetime signature $(+-)$, with a WZ term ${\cal L}_{\rm WZ}$ cubic in the axion fields,
and the scalar potential given by
\begin{align}
\label{eq:pot9}
  V_{\rm sugra} &= \tfrac12\, \rho^{5/9}\,  
  ( 2\,T^{IJ} \,T^{IJ} -T^{II} T^{JJ} ) \nonumber\\
&+\tfrac14 \, \rho^{-1/9}   \left( a^{IPQ} a^{KRS} \,T^{IK} T^{-1}_{PR} T^{-1}_{QS} - 2\, a^{IKP} a^{IKQ}\, T^{-1}_{PQ} \right)\nonumber\\
&  + \rho^{-13/9} \, Y_{IK} Y_{JL} T^{IJ} T^{KL}
+ {\cal O}(a^3)
\,,\ 
\end{align}
up to terms cubic in the axion and the auxiliary fields.

The two-dimensional theory (\ref{eq:gaugedLag}) has an ${SO}(9)$-invariant domain wall solution, which preserves half of the supersymmetries \cite{Ortiz:2012ib}. It is conveniently given in a frame  
\begin{equation}\label{eq:weyl}
    g_{\mu \nu} \longrightarrow \hat{g}_{\mu\nu}=\frac{25}{4}\,\rho^{-4/9}\, g_{\mu \nu} \,,
\end{equation}
in which the two-dimensional metric is an AdS$_2$ geometry
\begin{equation}
\mathrm{d}\hat{s}_2^2 = \frac1{r}\,\mathrm{d}t^2-\frac{1}{4\,r^2}\,\mathrm{d}r^2
\;,
\label{eq:dw1}
\end{equation}
of AdS radius $\ell_2=1$, and the scalar fields are given by
\begin{equation}
    T^{IJ} = \delta^{IJ}\;,\quad
    \rho=r^{-9/10}\;.
    \label{eq:dw2}
\end{equation}
This domain-wall solution has been used in \cite{Ortiz:2014aja} to compute holographic correlation functions in the strongly-coupled
BFSS matrix model. For the two-point functions of the operators dual to the ${\bf 44}$ and the ${\bf 84}$ scalars, one finds (in momentum space)
\begin{equation}
\label{eq:corr1}
\left\langle {\cal{O}}_{44} (0)\, {\cal{O}}_{44} (q) 
\right\rangle 
\propto q^{-6/5} \,,\qquad
\left\langle {\cal{O}}_{84} (0)\, {\cal{O}}_{84} (q) 
\right\rangle 
\propto q^{4/5} \,,
\end{equation}
respectively.
This agrees with the result obtained in
\cite{Hanada:2011fq} by Monte Carlo methods in the dual matrix quantum mechanics, but appears to be in tension  with the so-called quantum renormalisation group proposal \cite{Melo:2019zay}. The dual operators are identified as
\begin{equation}
{\cal O}_{44}  \propto  \text{tr} \big( \,{X}^{(\!(i} \,{ X}^{j)\!)} \big)\,,
\qquad
{\cal O}_{84}  \propto \text{tr} \big( [\,{ X}^i,\, \,{ X}^j] \,{ X}^k \big)~+~\mbox{fermions}
\,,
\end{equation}
in terms of the fundamental fields of the matrix quantum mechanics.

\subsection{Supersymmetric \texorpdfstring{$SO(p)$}{SO(p)}~\texttimes \texorpdfstring{$~SO(q)$}{SO(q)}  domain walls}

In this paper, we will be interested in supersymmetric solutions of the two-dimensional supergravity (\ref{eq:gaugedLag}) that are dual to particular deformations of the matrix model. 
Specifically, we will consider solutions that break the $SO(9)$ gauge symmetry down to $SO(p)\times SO(q)$, with $p+q=9$. Accordingly, the $SL(9)$ matrix $T^{IJ}$ will be of diagonal form, depending on one scalar field $x$ as
\begin{equation}\label{eq:Tdiag}
    T^{IJ} = 
    \begin{bmatrix}
    e^{x}\,\mathbb{I}_{p\times p} & 0 \\
    0 & e^{-px/q}\,\mathbb{I}_{q\times q}
    \end{bmatrix}
    \;.
\end{equation}
Moreover, we impose 
\begin{equation}
    A_\mu{}^{IJ} = 0 = Y^{IJ} = a^{IJK}
    \,,
\end{equation}
such that the Lagrangian (\ref{eq:gaugedLag}) reduces to
\begin{equation}
\label{eq:LagX}
\mathcal{L}_{\rm sugra} =
\sqrt{-g}\Big( -\rho R + 
\frac12 \,\rho \,K_p\,
\partial_\mu x \,\partial^\mu x 
- V_{p}(x)\Big)\,,
\end{equation}
with the normalization of the kinetic term, and the scalar potential given by
\begin{equation}
K_p = \frac{9\,p}{2\,q}  
\;,\qquad
V_p =-\frac12\,\rho^{5/9}
\left(
p\,(p-2)\,e^{2x}
+2pq\,e^{x(q-p)/q}
+q\,(q-2)\,e^{-2px/q}\right)
\;,
\label{eq:KV}
\end{equation}
respectively. 
We note the symmetry
\begin{equation}
    p \longleftrightarrow q\,,
    \qquad
    x \longleftrightarrow -\frac{p}{q}\,x
    \,,
    \label{eq:flipPQ}
\end{equation}
illustrating that $SO(p)\times SO(q)$ and $SO(q)\times SO(p)$
give rise to the same model upon however rescaling and 
flipping the sign of $x$,
which will become relevant in the later discussion.\footnote{
In this notation, the $SO(6)\times SO(3)$ domain wall of \cite{Ortiz:2014aja} is recovered 
with $p=6, q=3, x\rightarrow-x$, or equivalently with $p=3, q=6, x\rightarrow 2x$\,.
}
Let us also note that in the case $p=6$, $q=3$, there is also an $SO(6)\times SO(3)$ singlet among the axion fields transforming in the ${\bf 84}$ of $SO(9)$. In this case, one may thus consider more general domain wall solution as has been discussed in~\cite{Bobev:2024gqg}. Indeed, as explained in that reference, to make contact with the BMN deformation of the matrix model \cite{Berenstein:2002jq}, it is necessary to simultaneously turn on nontrivial profiles for $x$, $\rho$, and the axion singlet.

We will be mostly interested in supersymmetric solutions, i.e.\ solutions allowing for a non-vanishing Killing spinor. The full Killing spinor equations of (\ref{eq:gaugedLag}) have been given in \cite{Ortiz:2012ib}, corresponding to the vanishing of the variation of the gravitini, the dilatini and the 128 matter fermion fields. In the truncation (\ref{eq:Tdiag}), they take the form
\begin{equation}\label{eq:KS2}
    \begin{aligned}
0 &= \partial_{\mu}\epsilon
+\frac{1}{4}\,\omega_\mu{}^{\alpha \beta}\gamma_{\alpha \beta}\,\epsilon
+ \frac{7i}{36}\,\rho^{-2/9}\,T(x)\,\gamma_{\mu}\,\epsilon\,,\\
0 &= -\frac{i}{2}(\rho^{-1}\partial_{\mu}\rho)\,\gamma^{\mu}\,\epsilon
+\frac{1}{4}\rho^{-2/9}\,T(x)\,\epsilon\,,\\
0 &= (\partial_{\mu}x)\,\gamma^{\mu}\epsilon - \frac{i}{K_p}\rho^{-2/9}\,\partial_x T(x)\,\epsilon\,,
    \end{aligned}
    \end{equation}
for a spinor $\epsilon$ in the ${\bf 16}$ of $SO(9)$.
Here, $\omega_\mu{}^{\alpha \beta}$ is the spin connection and the $\gamma_\alpha$ denote the $SO(1, 1)$ gamma matrices. The function $T(x)$ is given by the trace of the matrix (\ref{eq:Tdiag})
\begin{equation}
    T(x) = T^{II}(x)=p\,e^x+q\,e^{-px/q}
    \,,
\label{eq:TS}
\end{equation}
and plays the role of a real superpotential.
In particular, the scalar potential (\ref{eq:KV}) can be cast into
the universal form
\begin{equation}
    V_p = -\frac{1}{18}\,\rho^{5/9}\left(7\,T^2-\frac{9}{K_p}\,(\partial_x T)^2\right).
    \label{eq:potentialT}
\end{equation}
In order to further simplify equations (\ref{eq:KS2}), we fix part of the diffeomorphism invariance and identify the scalar $x$ with the radial coordinate of the two-dimensional spacetime, assuming a locally invertible $x(r)$. With the diagonal ansatz
\begin{equation}\label{ds}
\mathrm{d}s_2^2 = \Tilde{f}(x)^2\mathrm{d}t^2-\Tilde{g}(x)^2\mathrm{d}x^2
\,,
\end{equation}
for the two-dimensional metric, and a potential of the general form (\ref{eq:potentialT}), 
the Killing spinor equations (\ref{eq:KS2}) reduce to the differential equations
\begin{equation}\label{eq:KS1}
\Tilde{g} = - \frac{K_p\, \rho^{2/9}}{\partial_x T} 
\,,\qquad
\Tilde{f}^{-1}\, \partial_x \Tilde{f}= - \frac{7\,K_p}{18}\frac{T}{\partial_x T}
\,,\qquad
\rho^{-1}\,\partial_x \rho  = -  \frac{K_p}{2}\, \frac{T}{\partial_x T} \,,
\end{equation}
with the associated Killing spinors satisfying
\begin{equation}\label{eq:KSeps}
   \gamma^1\epsilon = -i\epsilon
    \,.
\end{equation}
It is instructive to note that the equations (\ref{eq:KS1}) imply the 
full set of equations of motion obtained from variation of (\ref{eq:LagX}) for an arbitrary superpotential $T(x)$:
\begin{equation}
    \begin{aligned}
    0 =\;&  R - 
\frac12 \,K_p\,
\partial_\mu x \,\partial^\mu x 
+ \frac59\, \rho^{-1}\,V_{p}(x)\,,\\
    0 =\;&  \nabla_\mu \partial^\mu\rho -  V_{p}(x)\,,\\
    0 =\;&
    K_p\,
\nabla_\mu ( \,\rho\,\partial^\mu x )
- \partial_x V_{p}
    \,.\\
    \end{aligned}
    \label{eq:eom}
\end{equation}
For example, the Ricci scalar for the metric (\ref{ds}) is given by
\begin{equation}
    R = \frac{2}{\tilde f\tilde g}\,\partial_x \big(\tilde{f}' \tilde{g}^{-1}\big) 
    =\frac{7}{162}\,\rho^{-4/9}\,\Big(5\,T^2-\frac{18}{K_p}\,(\partial_x T)^2 \Big)
    \;,
    \label{eq:RicciScalar}
\end{equation}
where we have used (\ref{eq:KS1}) for the second equation. This in turn shows the first field equation in (\ref{eq:eom}). With $T$ explicitly given by (\ref{eq:TS}), the differential equations (\ref{eq:KS1}) can be integrated to
\begin{equation}
\rho(x)=\left(e^{px/q}-e^{-x}\right)^{-9/4}
\,,
\label{eq:solution1}
\end{equation}
for the dilaton, and
\begin{equation}
\tilde{f}(x)=\rho(x)^{7/9}
\,,\quad
\tilde{g}(x) = - \frac{9\, \rho(x)^{2/9}}{2q}\,
\left(e^{x}-e^{-px/q}\right)^{-1}
\,,
\label{eq:solution2}
\end{equation}
for the components of the two-dimensional metric. 
Here and in the following, without loss of generality, we will restrict to the region $x>0$. At $x=0$, the two-dimensional metric has a coordinate singularity, and the dilaton vanishes.
The region $x<0$ is covered by the same analysis via the symmetry (\ref{eq:flipPQ}), i.e.\ upon exchanging $p$ and $q$.
The Killing spinor of (\ref{eq:KS2}) is finally given by
\begin{equation}\label{eq:KSeps1}
    \epsilon(x) = a(x) \epsilon_0 ~~\textrm{with}~~\gamma^1\epsilon_0 = -i\epsilon_0
    \,,
\end{equation}
with constant $\epsilon_0$, and 
a function $a(x)$ that is obtained from integrating the first equation of (\ref{eq:KS2}). 
Moreover, equation (\ref{eq:KSeps}) shows that all these solutions preserve half of the supersymmetries. As a solution of the maximal supergravity (\ref{eq:gaugedLag}), they hence preserve sixteen supercharges.%

For the later holographic analysis, 
it will be convenient to change the two-dimensional frame by the Weyl rescaling (\ref{eq:weyl})
\begin{equation}\label{eq:weyl1}
    g_{\mu \nu} \rightarrow \hat{g}_{\mu\nu}=\rho^{-4/9}\, g_{\mu \nu} \,,
\end{equation}
and a simultaneous change of coordinates, $x = r^{2/5}$.
We then recover the metric of an asymptotically AdS spacetime coupled to a dilaton given by
\begin{align}
    &\rd \hat{s}^2_2 = \hat{f}(r)^2 \rd t^2 - \hat{g}(r)^2\rd r^2 \,,\nonumber\\
    &\hat{f}(r) = \left(e^{px/q}-e^{-x}\right)^{-5/4}
    \,,\quad
    \hat{g}(r)=\frac{9}{5\,q}\,x^{-3/2}\left(e^{x}-e^{-px/q}\right)^{-1}\,.
    \label{dsweyl63}
\end{align}
whose near-horizon expansion
\begin{equation}\label{AdS263}
\begin{split}
   & \rd \hat{s}_2^2 \; \underset{r \rightarrow 0}{\sim}\; \left(
    \frac{\rd t^2}{r} - \frac{\rd r^2}{4\,r^2}\right)\left(1+{\cal O}(r^{2/5})\right) \,,\\[1ex]
    &\rho \underset{r \rightarrow 0}{\sim} r^{-9/10}\,\left(1+{\cal O}(r^{2/5})\right) 
    \,,\qquad x   \; \underset{r \rightarrow 0}{\sim}\; {\cal O}(r^{2/5})
    \,,
    \end{split}
\end{equation}
approaches the ${\rm SO}(9)$-symmetric domain wall (\ref{eq:dw1}), (\ref{eq:dw2}), 
after rescaling of $t$ by a numerical factor.
In particular, we find that all the above domain wall solutions share the same asymptotic behavior.
Following the analysis of \cite{Ortiz:2014aja}, taking into account the correlators of the undeformed theory (\ref{eq:corr1}), one may deduce that the holographic dual of this domain wall solution is a VEV deformation of the BFSS matrix model just as the SO$(3)\times$ \!SO(6) solution of \cite{Ortiz:2014aja}, as further discussed in Appendix \ref{sec:appendix}. This is also in agreement with the recent discussion in \cite{Bobev:2024gqg}.

\subsection{Higher-dimensional uplift}

The domain wall solution (\ref{eq:solution1}), (\ref{eq:solution2}) can be uplifted to ten- and eleven-dimensional supergravity. As the ansatz (\ref{eq:Tdiag}) lives within the $U(1)^4$ truncation of the two-dimensional model, the uplift to IIA supergravity can be obtained from the uplift formulas given by \cite{Anabalon:2013zka}. More general, the embedding of the full model (\ref{eq:gaugedLag}) is given by the general formulas derived in~\cite{Bossard:2023jid}.

In $D=11$ solution, this gives rise to a purely geometrical solution (i.e.\ with vanishing three-form) with the metric given by
\begin{align}\label{uplift11dpq}
    \rd s^2_{11} = &-2\, \rd t \rd z - \frac{(e^{9x/q}-1)^{7/2}}{(1-\mu^2)\,e^{9\,(7-p)\,x/(2q)}+\mu^2\,e^{9x/2}}\,\rd z^2 
    - \frac{81\,e^{9x/q}\Big(1+\mu^2(e^{9x/q}-1)\Big)}{4q^2(e^{9x/q}-1)^{3}} \,\rd x^2
    \nonumber\\[1ex] 
    &
       -\frac{\mu^2}{e^{9x/q}-1}\,\rd \Omega^2_{p-1}  
    -\frac{1-\mu^2}{1-e^{-9x/q}}\,\rd \Omega^2_{q-1} 
  - \frac{(1-\mu^2)+\mu^2\,e^{9x/q}}{(1-\mu^2)(e^{9x/q}-1)}\,\rd \mu^2 
    \,,
\end{align}
in terms of coordinates $\{t, z, x, \mu\}$
together with the coordinates of two round 
spheres $S^{p-1}$ and $S^{q-1}$.

With respect to the symmetry (\ref{eq:flipPQ}) of the two-dimensional model, we observe that the metric (\ref{uplift11dpq}) is indeed invariant under this symmetry upon further exchange $\mu^2\rightarrow1-\mu^2$, however up to imaginary numerical factors appearing within the metric components. This reflects the fact that the two-dimensional solution (\ref{eq:solution1}), (\ref{eq:solution2}) is only well-defined for positive $x$ and turns complex for a naive analytic continuation. Consequently, the family (\ref{uplift11dpq})  of $D=11$ solutions describes eight distinct solutions for the values $p=1, 2, \dots, 8$.

Eventually, the $D=11$ metric can be considerably simplified by the following coordinate transformations
\begin{equation}
\label{coord11dpq}
    r_p^2 = \frac{\mu^2}{e^{9x/q}-1} \,,\qquad
        r_q^2 = \frac{e^{9x/q}(1-\mu^2)}{e^{9x/q}-1}
        \,,\qquad
    x^{\pm} = t \pm (t + z)\,,
\end{equation}
upon which the metric (\ref{uplift11dpq}) takes the familiar form of a pp-wave solution
\cite{Gauntlett:2002cs}
\begin{equation}\label{metric11dpq}
    \rd s_{11}^2 =  \rd x^- \rd x^+ - H(r_p,r_q)\,(\rd x^-)^2 - \Big(\rd r_p^2 + r_p^2 \,\rd \Omega_{p-1}^2 + \rd r_{q}^2 + r_q^2\,\rd \Omega_{q-1}^2 \Big)
    \,,
\end{equation}
with the function $H(r_p,r_q)$ given by
\begin{align}
\label{eq:Hpq}
    H(r_p,r_q) &=  -1+
    2^{(p-7)/2}\,r_q^{2-p}\,\gamma^{-1}
    \left(1+r_p^2+r_q^2-\gamma\right)^{5/2}
    \left(-1-r_p^2+r_q^2+\gamma\right)^{(2-p)/2}
    \,,\\[2ex]
    \gamma &\equiv \sqrt{(1+r_p^2+r_q^2+2r_q)(1+r_p^2+r_q^2-2r_q)}
    \,.
\end{align}
As required for consistency, one may verify that the function $H(r_p,r_q)$ satisfies the Laplace equation 
\begin{equation}
\Delta H = 
r_p^{1-p} \,\partial_{r_p} \left(
r_p^{p-1}\,\partial_{r_p} H \right)
+r_q^{1-q} \,\partial_{r_q} \left(
r_q^{q-1}\,\partial_{r_q} H \right)
= 0
\,,
\end{equation}
on the Euclidean space $\mathds{E}^9$.
From the eleven-dimensional perspective, the metric (\ref{metric11dpq}) thus is part of a well understood class of supersymmetric solutions. The non-trivial result of our analysis is the fact that these particular pp-wave solutions live within a consistent truncation to two-dimensional supergravity. In turn, this allows us to perform certain holographic computations such as the dynamics of fluctuations around the backgrounds (\ref{metric11dpq}) within the framework of the two-dimensional theory. The solution for $p=6$ (and $q=3$) was already constructed in \cite{Ortiz:2014aja}, here we have generalized the construction to the full family of solutions with $p\in\{1, \dots, 8\}$.

\subsection{Brane interpretation}

From the ten-dimensional point of view the solution can in fact be interpreted as the near-horizon limit of a distribution of D0-branes with $SO(p)\times SO(q)$ symmetry, similarly to the multi-centered solutions of \cite{Freedman:1999gk} for D3-branes. 
To make the brane distribution explicit,  let us consider the limit, 
\begin{equation}\label{H11dpqdev}
    H(r_p,r_q) \; \underset{r_p \rightarrow 0}{\sim}\;
    r_p^{2-p}\,(1-r_q^2)^{(p-4)/2} + \mathcal{O} \left(r_p^{4-p}\right)\,.
\end{equation}
This suggests that the D0-branes are localized at $r_p = 0$, in $p$ of the nine transverse dimensions. We can check 
this  directly, by comparing to  the canonical  $SO(p)\times SO(q)$-invariant harmonic function $h(r_p , r_q)$ in $\mathds{R}^9$, given by, 
\begin{equation}\label{Hsigmapq}
h(r_p,r_{9-p}):= \int_{\mathbb{R}^9}\d^9\! y ~\!\delta(\vec{y}_p) ~\! \sigma_p(|\vec{y}_{9-p}|) \frac{1}{|\vec{r}_9-\vec{y}_9|^7}
~,
\end{equation}
where we have set $\vec{r}_9=\vec{r}_p+\vec{r}_{9-p}$, $r_p=|\vec{r}_p|$, $r_{9-p}=|\vec{r}_{9-p}|$. 
The   harmonic function $h(r_p , r_q)$ above describes  D0-branes
 localized at $\vec{y}_p=0$ and smeared in the remaining  $(9-p)$ transverse directions, with charge density $\sigma_p(|\vec{y}_{9-p}|)$.  
 
In order to compare $H(r_p,r_q)$  to $h(r_p,r_q)$,  
it is useful to rewrite the latter 
using spherical coordinates in $\mathbb{R}^{9-p}$, choosing the ``z-axis''  to be along $\vec{r}_{9-p}$, so that, for  $1\leq p\leq7$, (\ref{Hsigmapq}) takes the form, 
\begin{equation}\label{Hsigmadevpq}
h(r_p,r_{9-p})={S}_{7-p} \int_0^\infty \d\rho~\!\rho^{8-p}\sigma_p(\rho)\int_{-1}^1\d x~\!(1-x^2)^{\frac{6-p}{2}} (r_{p}^2+r_{9-p}^2+\rho^2-2x\rho~\! {r}_{9-p})^{-\frac72}
~,
\end{equation}
where ${S}_{n-1}={2\pi^{\frac{n}{2}}}/{\Gamma(\frac{n}{2})}$ is the area of the $n$-dimensional unit sphere.  
For $p=8$,   (\ref{Hsigmapq})  takes the form,
\eq{\label{3jhg}
h(r_8,r_{1})=2\pi
\int_{\mathbb{R}} \d y ~\!\sigma_8(y)
(r_{8}^2+r_{1}^2+y^2-2y~\! {r}_{1})^{-\frac72}
~,}
Comparing, order by order, the series expansions of  the harmonic function  $H(r_p,r_q)$  in (\ref{eq:Hpq})  and  $h(r_p,r_q)$  in \eqref{Hsigmadevpq}, \eqref{3jhg}, we find exact agreement, 
provided the charge  distribution of the smeared branes in the $(9-p)$ transverse directions is given by  Table \ref{tab:branes_all_p}.  
Specifically, the two harmonic functions are related by, 
\eq{\label{238gbfkj}
H(r_p,r_q)=-1+C_p\,h(r_p,r_q)
~,}
for  certain numerical coefficients $C_p$ depending on $p$.  
\renewcommand{\arraystretch}{2}
\begin{table}[H]
\begin{center}
\centering
\begin{tabular}{| c || c | c | }
\hline
\cellcolor[gray]{0.9} $SO(p)\times SO(q)$ & \cellcolor[gray]{0.9} p odd &  \cellcolor[gray]{0.9} p even \\[4pt]
\hline
  $SO(2)\times SO(7)$  &  $\sigma_7({y}) = (1-{y}^2)^{3/2}$ & $\sigma_2({y}) = \delta (1-{y})$ \\[7pt]
\hline
 $SO(4)\times SO(5)$ &  $\sigma_5({y}) = \sqrt{1-{y}^2}$  &  $\sigma_4({y}) = 1$  \\[7pt]
\hline
 $SO(6)\times SO(3)$ & $\sigma_3({y}) = \frac{1}{\sqrt{1-{y}^2}}$   & $\sigma_6({y}) = 1 - {y}^2$  \\[7pt]
\hline
   $SO(8)\times SO(1)$  &   $\sigma_1({y}) = \left[\frac{1}{(1-{y}^2)^{3/2}}\right]_\text{reg}$  & $\sigma_8(y) = (1-y^2)^2$   \\[7pt]
\hline
\end{tabular}
\end{center}
\caption {Charge  distribution $\sigma_p(y)$, $|y|\leq 1$, of the D0 branes, smeared along $(9-p)$ of the transverse directions; 
 for $|y|>1$, $\sigma_p\equiv 0$. For $p=1$, the corresponding charge  distribution is appropriately regularized, {\it cf}.~eq.~\eqref{08}  of the main text. For $p=3$, we recover the result of \cite{Ortiz:2014aja}.}
    \label{tab:branes_all_p}
\end{table}
\renewcommand{\arraystretch}{1}
We see that for all $p>2$, 
 the D0-brane charge distribution is given by, 
\begin{equation}\label{sigma_p}
    \sigma_p(y) = \left\{
\begin{array}{ccc}
(1-y^2)^{\frac{p-4}{2}}&,&|y|\leq1\\
0&,&|y|>1
\end{array}
\right.  ~.
\end{equation}
 For $p=1$, the integral obtained using the formula \eqref{sigma_p} is divergent and needs to be regularized. 
The correct regularization is achieved by integrating by parts the  formula obtained from \eqref{Hsigmadevpq} by using the ``bare'' charge distribution $ {(1-{y}^2)^{-3/2}}$, 
\eq{
h(r_1,r_{8})=-{S}_{6}  \int_{-1}^1\d x~\!(1-x^2)^{\frac{5}{2}}  \int_0^1 \d\rho~\!  \frac{1}{\sqrt{1-{\rho}^2}}~\!\frac{\d}{\d\rho}\left[\rho^{6}(r_{1}^2+r_{8}^2+\rho^2-2x\rho~\! {r}_{8})^{-\frac72}\right]
~,
}
while dropping the (divergent) boundary term. This amounts to setting, 
\eq{\label{08}
\sigma_1({y}) = \left[\frac{1}{(1-{y}^2)^{3/2}}\right]_\text{reg}:=  \frac{1}{(1-{y}^2)^{3/2}} -\frac{\delta(1-y)}{\sqrt{1-{y}^2}}
~.}
The delta-function charge distribution for the case $p=2$ of Table \ref{tab:branes_all_p}, is analogous to the D3-brane charge distribution of  Table 1 of \cite{Freedman:1999gk}.

\section{Lagrangian for scalar fluctuations}

Our goal in this paper is the computation of holographic correlation functions around the supersymmetric $SO(p) \times SO(q)$ domain wall backgrounds constructed in (\ref{eq:solution1}), (\ref{eq:solution2}) above. 
In a first step, we will compute the effective action quadratic in the scalar fluctuations around these background solutions.

As mentioned above, for the holographic analysis it is convenient to change the two-dimensional frame by the Weyl rescaling (\ref{eq:weyl1}) and to rotate to Euclidean signature. The resulting background Lagrangian (\ref{eq:LagX}) takes the form
\begin{align}
\label{eq:LagXWeyl}
\mathcal{L}_{\rm sugra} =
\sqrt{-g}\,\rho\,& \Big( R 
+\frac49\,\rho^{-2}\partial_\mu\rho\, \partial^\mu \rho
-\frac{9\,p}{4\,q} \,
\partial_\mu x \,\partial^\mu x 
\nonumber\\
&{}
+ \frac12
\left(
p\,(p-2)\,e^{2x}
+2pq\,e^{x(q-p)/q}
+q\,(q-2)\,e^{-2px/q}\right)
\Big)\,,
\end{align}
while the background solution (\ref{eq:solution1}), (\ref{eq:solution2}) becomes
\begin{align}\label{eq:background}
    &\rd {s}^2 = \left(e^{px/q}-e^{-x}\right)^{-5/2} \rd t^2 
    + \frac{81}{4 q^2}\left(e^{x}-e^{-px/q}\right)^{-2} \rd x^2 \,,\qquad    
    \rho=\left(e^{px/q}-e^{-x}\right)^{-9/4}
    \,.
\end{align}

The scalar field fluctuations around this background are computed by expanding the Lagrangian (\ref{eq:gaugedLag}) to quadratic order in the components of the $SL(9)$ matrix $T^{IJ}$ and the axions $a^{IJK}$. Since the background breaks the gauge symmetry down to
\begin{equation}
    SO(p) \times SO(q) \,,
    \label{eq:SOpq}
\end{equation}
the $\textbf{44} + \textbf{84}$ scalar fields of the theory organize in representations of (\ref{eq:SOpq}) as
\begin{align}
 T^{IJ}\,:\;\;   \mathbf{44} \,&\rightarrow (\tinyyoung{&\cr}\;,1) \oplus (\tinyyoung{\cr}\;,\tinyyoung{
\cr}\,) \oplus (1,\tinyyoung{&\cr}\,)   \oplus (1,1)\,,
\nonumber\\
 a^{IJK} \,:\;\;   \mathbf{84} \,&\rightarrow (\tinyyoung{\cr\cr\cr}\;,1) \oplus (\tinyyoung{\cr\cr}\;,\tinyyoung{\cr}\,) \oplus (\tinyyoung{\cr}\;,\tinyyoung{\cr\cr}\,) \oplus (1,\tinyyoung{\cr\cr\cr}\,)
    \,,
    \label{eq:4484topq}
\end{align}
according to the distribution of their indices over the two factors of (\ref{eq:SOpq}).
Fluctuations of the $(1,1)$ singlet scalar exhibit more complicated couplings, as they mix with the fluctuations of the dilaton $\rho$ and the two-dimensional metric. We will discuss them below. 
For the other scalar fluctuations, their quadratic Lagrangian takes the universal form
\begin{equation}
\mathcal{L}_{\phi} =
-\frac12\,\sqrt{-g}\,\Big(  
K_\phi \,\partial_\mu \phi \,\partial^\mu \phi  + 
M_\phi\,\phi^2
\Big)
\,,
\label{eq:LflucPhi}
\end{equation}
of a kinetic and a mass term, with normalization constants $K_\phi$, $M_\phi$,
since to quadratic order these fluctuations can only couple to themselves.

For the scalars in the ${\bf 44}$, their fluctuations are conveniently described by representing the matrix $T^{IJ}={\cal V}^I{}_A{\cal V}^J{}_A$ in terms of an $SL(9)/SO(9)$ coset representative ${\cal V}$ which carries the scalar fluctuations according to
\begin{equation}\label{eq:vvback}
    \mathcal{V} = \mathring{\mathcal{V}}_{\rm background}\Big(\mathbb{I}_{9x9} + X +\frac{1}{2}X^2 + ...\Big)
    \,.
\end{equation}
With $\mathring{\mathcal{V}}_{\rm background}$ corresponding to the background  (\ref{eq:Tdiag}), plugging (\ref{eq:vvback}) into the Lagrangian (\ref{eq:gaugedLag}) and expanding to quadratic order in the fluctuations, the couplings take the form (\ref{eq:LflucPhi}) with the coefficients $\{K_\phi, M_\phi\}$ respectively given by
\begin{align}
\begin{array}{rll}
(\tinyyoung{&\cr}\;,1) :&  \ K=\rho\,, &\ 
M=\rho\,\Big(   (8-2p)\,e^{2x}  -2\,q\, e^{x - px/q} \Big)\,, \\
(\tinyyoung{\cr}\;,\tinyyoung{\cr}\,) :&\ K=\rho\,, &\ M=\rho\,\Big(   (3-p)\,e^{2x}  +  (3-q)\,e^{-2 p x/q}  -7\, e^{x - px/q} \Big)\,, \\
(1,\tinyyoung{&\cr}\,) :&\ K=\rho\,, &\ M= \rho\,\Big(  (8-2q)\,  e^{-2 p x/q}-2\,p\, e^{x - px/q} \Big) \,,
\end{array}
\label{eq:KM44}
\end{align}
for the different components of (\ref{eq:4484topq}). 
For the mass term $M$ of the $(\tinyyoung{\cr}\;,\tinyyoung{\cr}\,)$,
one has to take into account an additional contribution originating from the kinetic term $\partial_\mu x \partial^\mu x$ upon using the background solution (\ref{eq:solution2})
\begin{equation}
\frac{81}{8\,q^2}\,\rho\,\partial_\mu x\partial^\mu x = 
\frac{81}{8\,q^2}\,\rho\,g^{xx} 
=-\frac12\,\rho\,
(e^x-e^{-px/q})^2
\,.
\end{equation}

For the scalars in the ${\bf 84}$, the computation is even more straightforward, upon expanding the axion terms of the Lagrangian (\ref{eq:gaugedLag}) on the background (\ref{eq:Tdiag}). The resulting couplings are again of the form (\ref{eq:LflucPhi}) with
\begin{align}
\begin{array}{rll}
(\tinyyoung{\cr\cr\cr}\;,1) :&\ K=\rho^{1/3}\,e^{-3x}\,, &\ M=3\,\rho^{1/3}\,e^{-x}\,, \\
(\tinyyoung{\cr\cr}\;,\tinyyoung{\cr}\,) :&\ K=\rho^{1/3}\,e^{-2x+px/q} \,,&\ M=\rho^{1/3}\,\Big(4\,e^{-x}-e^{-2x-px/q}\Big)\,, \\
(\tinyyoung{\cr}\;,\tinyyoung{\cr\cr}\,) :&\ K=\rho^{1/3}\,e^{-x+2px/q} \,,&\ M=\rho^{1/3}\,\Big(4\,e^{px/q}-e^{x+2px/q}\Big) \,,\\
(1,\tinyyoung{\cr\cr\cr}\,) :&\ K=\rho^{1/3}\,e^{3px/q}\,, &\ M=3\,\rho^{1/3}\,e^{px/q} \,.
\end{array}
\label{eq:KM88}
\end{align}

\section{Holographic computation of correlators}

In this section, we review and extend the results of \cite{Ortiz:2014aja} to set up the holographic renormalization and the computation of holographic two-point functions for the example of the $SO(3) \times SO(6)$ domain wall. According to (\ref{eq:4484topq}), the scalars around the $SO(3) \times SO(6)$ domain wall break into the representations
\begin{align}
 T^{IJ}\,:\;\;   \mathbf{44} \,&\rightarrow (5,1) \oplus (3,6) \oplus (1,20)   \oplus (1,1)\,.
\nonumber\\
 a^{IJK} \,:\;\;   \mathbf{84} \,&\rightarrow (1,1) \oplus (3,6) \oplus (3,15) \oplus (1,20)\,
    \,,
    \label{eq:4484to36}
\end{align}
of $SO(3)\times SO(6)$.
We will first discuss the fluctuations from the $\mathbf{44}$ in the $(5,1) \oplus (3,6) \oplus (1,20)$ whose action is given by (\ref{eq:LflucPhi}), (\ref{eq:KM44}). We then discuss separately the fluctuation of the `active' scalar $(1,1)\in{\bf 44}$ whose fluctuations couple to the fluctuations of the dilaton and the two-dimensional metric. Finally, we discuss the case of the fluctuations in the ${\bf 84}$ whose actions are given by (\ref{eq:LflucPhi}), (\ref{eq:KM88}).

In the next section, we then generalize the construction to all the supersymmetric $SO(p)\times SO(q)$ domain walls.

\subsection{On-shell action, renormalization, and correlators for scalars in the \textbf{44} sector}
\label{subsec:2pt44}

In this subsection, we focus on setting up the holographic renormalization for scalars in the \textbf{44} representation, and the computation of their two-point correlation functions. 
We follow the general scheme of \cite{Bianchi:2001de,Bianchi:2001kw,Kanitscheider:2008kd}, see also \cite{Cvetic:2016eiv} specifically for the case of a two-dimensional bulk.
After Weyl rescaling, the relevant Euclidean action is given by the sum of (\ref{eq:LagXWeyl}) and (\ref{eq:LflucPhi}), (\ref{eq:KM44}) for the different representations. In order to facilitate the comparison with \cite{Ortiz:2014aja}, we match the earlier conventions by rescaling the background scalar in this section as $x\rightarrow2x$.

\subsubsection{Asymptotic analysis}

The equations of motion derived from (\ref{eq:LagXWeyl}), (\ref{eq:LflucPhi}), (\ref{eq:KM44}), are given by
\begin{equation}\label{EOMperturb63}
    \begin{split}
        0 = ~&\rho^{-1}\nabla \partial \rho - \frac{3}{2}e^{-2x}(8 + 12e^{3x} + e^{6x}) - F_i(x)\,x_i^2 ~,\\
        0 = ~&\rho^{-1}(\nabla_{\mu}\partial_{\nu}\rho - \frac{1}{2}g_{\mu\nu}\nabla \partial \rho) -\frac{4}{9}\rho^{-2}(\partial_{\mu}\rho\partial_{\nu}\rho - \frac{1}{2}g_{\mu\nu}(\partial \rho)^2) \\ & + \frac{9}{2}(\partial_{\mu}x\partial_{\nu}x - \frac{1}{2}g_{\mu\nu}(\partial x)^2) +  2(\partial_{\mu} x_i\partial_{\nu}x_i - \frac{1}{2}g_{\mu\nu}(\partial x_i)^2)\,,\\
        0 = ~&R + \frac{4}{9}\rho^{-2}(\partial \rho)^2 - \frac{8}{9}\rho^{-1}\nabla\partial\rho -\frac{9}{2}(\partial x)^2+\frac{3}{2}e^{-2x}(8 + 12e^{3x} + e^{6x}) - 2((\partial x_i)^2 - \frac{1}{2}F_i(x)x_i^2 )\,,\\
        0 =  ~&\rho^{-1}\nabla (\rho \partial x) - \frac{2}{3}e^{-2x}(4 - 3e^{3x} - e^{6x}) +\frac{1}{9} F_i'(x) x_i \,\\
        0 = ~&\rho^{-1}\nabla (\rho \partial x_i) + \frac{1}{2} F_i(x) x_i\,,
    \end{split}
\end{equation}
with $i \in {(5,1), (1,20), (3,6)}$, labelling the different representations, and the scalar functions
\begin{equation}\label{Fperturb63}
    F_{(5,1)} = - 4 e^x(e^{3x}-6)~,~~ F_{(1,20)} = 4 (2e^{-2x}+3e^{x}) ~,~~ F_{(3,6)} = 2 e^{-2x}(3 + 7e^{3x} )~,~~
\end{equation}
found from (\ref{eq:KM44}), which capture the interactions of the scalar fluctuations with the background $x(t,r)$ from (\ref{dsweyl63}).

As a second step, similarly to the procedure implemented in \cite{Ortiz:2014aja}, the effective action is most conveniently evaluated on-shell using the dilaton field equation. This then leads to a contribution located at the boundary of the asymptotically AdS spacetime background (\ref{dsweyl63}),
\begin{equation}\label{Sonshell63}
S_{\rm on-shell} = \frac{1}{2}\int_{r=\epsilon} \rd t \sqrt{|h|}\,\Big(\frac{4}{9}n^{\mu}\partial_{\mu}\rho + \rho K\Big)    \,,
\end{equation}
with the integration performed on the regularizing surface $r=\epsilon$ for small $\epsilon$. Here, $h$ denotes the induced metric on the boundary, $n^{\mu}$ is the normal vector to that surface and $K= \nabla_{\mu} n^{\mu}$ is the extrinsic curvature of the boundary. In the following, we can deal with the different fluctuations separately since at the quadratic level there is no mixing among the different representations. We parametrize the scalar fluctuations as
\begin{equation}
x_{i,{\rm fluc}} (t,r) =  \sum_{k\ge2} c_{x_i,k}(t)\, r^{k/5}\,,\quad 
\mathrm{for}~i \in (5,1), (1,20), (3,6)\,,
\label{eq:seriesXi}
\end{equation}
where the leading $r$-power is extracted from the field equations, i.e.\ the last equation in (\ref{EOMperturb63}).
Evaluation of the on-shell action on the background shows that the dilaton and extrinsic curvature terms diverge as
\begin{equation}\label{estimationsonshell63}
    \sqrt{|h|}n^{\mu}\partial_{\mu}\rho  \underset{r \rightarrow 0}{\sim}  r^{-7/5} ~~,~~ \sqrt{|h|}\rho K  \underset{r \rightarrow 0}{\sim} r^{-7/5}
    \,.
\end{equation}
In the following analysis, we will thus consider the series expansion (\ref{eq:seriesXi}) only up to the power $k=7$\,.
For the moment, we do not put a source in the dilaton-gravity sector, i.e.\ we expand the fields in this sector as
\begin{equation}\label{frhoback63}
  \begin{split}
        f(t,r) &= f_b(r)(1+f_{\rm fluc}(t,r)) \,,\quad \rho(t,r) = \rho_b(r)(1+\rho_{{\rm fluc}}(t,r))\,,  \\ 
        g(t,r) &= g_b(r)(1+g_{\rm fluc}(t,r)) \,,\quad x(t,r) = x_b(r)(1+x_{{\rm fluc}}(t,r))\,,
\end{split}  
\end{equation}
where $f_b$, $x_b$, $g_b$ and $\rho_b$ denote the background (\ref{dsweyl63}) and their fluctuations $f_{\rm fluc}(t, r)$, $g_{\rm fluc}(t, r)$, $x_{\rm fluc}(t, r)$, and $\rho_{\rm fluc} (t, r)$ are determined as  the backreaction of the scalar fluctuations. 
Explicitly, we parametrize them as 
\begin{equation}\label{frhoback63v2}
        f_{{\rm fluc}}(t,r) =  \sum_{k\ge4} c_{f,k}(t)\, r^{k/5} \,,\quad 
        \rho_{{\rm fluc}}(t,r)=  \sum_{k\ge4} c_{\rho,k}(t)\, r^{k/5} \,,  \quad 
        x_{{\rm fluc}}(t,r)=  \sum_{k\ge2} c_{x,k}(t)\, r^{k/5} \,,
\end{equation}
while we use the remaining freedom of diffeomorphisms in order to eliminate the fluctuations $g_{fluc}(t,r)$. Plugging these expansions into the field equations (\ref{EOMperturb63}), we find that the free coefficients in the scalar fluctuations (\ref{eq:seriesXi}) are given by $c_{x_i,2}(t)$ and $c_{x_i,5}(t)$. In turn, all other coefficients in the series expansions (\ref{EOMperturb63}), (\ref{frhoback63v2}) are determined as functions thereof.
Explicitly, the non-vanishing higher coefficients are given by:
\begin{equation}\label{exprcoeffdev63relation}
    \begin{split}
 c_{x,2} &= a_{i,2} \,c_{x_i,2}^2 ~,~ c_{x,4}=a_{i,4}\, c_{x_i,2}^2 ~,~ c_{x,5}=-\tfrac{783}{1080}c_{x,3} - a_{i,5}\,c_{x_i,2}c_{x_i,5}\,, \\
c_{x,6} &=a_{i,6,1}\, c_{x_i,2}^2 + a_{i,6,2} \,c_{x_i,2}^4 +a_{i,j,6,3}\,c^2_{x_i,2}c^2_{x_j,2} \,,\\ 
c_{x,7} &=  -\tfrac{29627}{53760}c_{x,3} + a_{i,7,1} \,c_{x_i,2}^2c_{x,3} + a_{i,7,2}\,c_{x_i,2}c_{x_i,5}  + a_{i,7,3}\,c_{x_i,2}\Ddot{c}_{x_i,2} +\tfrac{719}{1440}c_{f,7} \,,\\[1ex]
c_{f,4} &= b_{i,4}\, c_{x_i,2}^2 \,,\quad c_{f,6}=b_{i,6}\, c_{x_i,2}^2 \,,\\[1ex]
c_{\rho,4}&= d_{i,4} \,c_{x_i,2}^2 ~,~ c_{\rho,6} = d_{i,6}\, c_{x_i,2}^2 \,,\quad c_{\rho,7}=- \tfrac{7200}{7}c_{x,3}-\tfrac{800}{21}c_{x_i,2}c_{x_i,5} -720c_{f,7}  \,,\\[1ex]
c_{x_i,4}&=h_{i,4}\,c_{x_i,2} \,,\quad c_{x_i,6} = h_{i,6,1}\,c_{x_i,2}^3+ h_{i,6,2}\,c_{x_i,2} + h_{i,j \neq i,6,3}\,c_{x_i,2}c_{x_j,2}^2 \,,\\ 
c_{x_i,7}&=h_{i,7,1}\,c_{x_i,2}c_{x,3} + h_{i,7,2}\,c_{x_i,5} + h_{i,7,3}\,\Ddot{ c}_{x_i,2}
\,,
    \end{split}
\end{equation}
with $c_{f,7}(t)$, $c_{x,3}(t)$, $c_{x_i,2}(t)$ and $c_{x_i,5}(t)$ still left unconstrained, and dots representing time derivatives. The numerical coefficients appearing in these expansion, are summarized in the following tables
\renewcommand{\arraystretch}{1.5}
 \begin{table}[H]
    \centering
    \begin{tabular}{|c||c|c|c|c|c|c|c|c|c|}
         \hline
         \cellcolor[gray]{0.9} & \cellcolor[gray]{0.9} $a_2$ & \cellcolor[gray]{0.9} $a_4$ & \cellcolor[gray]{0.9} $a_5$&\cellcolor[gray]{0.9} $a_{6,1}$ &\cellcolor[gray]{0.9} $a_{6,2}$ &\cellcolor[gray]{0.9} $a_{7,1}$& \cellcolor[gray]{0.9}$a_{7,2}$ & \cellcolor[gray]{0.9}$a_{7,3}$ \\ 
         \hhline {|=||=|=|=|=|=|=|=|=|}
        \cellcolor[gray]{0.9} (5,1) & $\frac{4}{9}$ & $\frac{520}{774}$ & $-\frac{2}{270}$ & $\frac{12822139109}{16046598960}$
         & $\frac{6565}{31347}$ & $\frac{99742}{609525}$ & $\frac{55381}{1814400}$  & $-\frac{3599}{113400}$  \\
         \hline
        \cellcolor[gray]{0.9} (1,20) & $-\frac{2}{9}$ & $-\frac{211}{774}$ & $\frac{1}{270}$ & $\frac{56210924513}{96279593760}$ & $\frac{379}{156735}$ & $\frac{202543}{9752400}$ & $\frac{19141}{1209600}$  & $ \frac{3599}{226800}$  \\
         \hline
        \cellcolor[gray]{0.9} (3,6) & $\frac{1}{9}$ & $-\frac{555}{774}$ & $-\frac{1}{540}$ & $-\frac{200557561}{1188636960}$ & $\frac{139}{167184}$ & $\frac{810043}{39009600}$ & $\frac{12197}{1036800}$ & $-\frac{3599}{453600}$ \\
         \hline 
    \end{tabular}
\label{numericalcoeffdev63v1}
\end{table}
\begin{table}[H]
    \centering
    \hfill
    \begin{tabular}{|c||c|c|c|}
     \hline
     \cellcolor[gray]{0.9} $a_{i,j,6,3}$    & \cellcolor[gray]{0.9} (5,1) & \cellcolor[gray]{0.9} (1,20) & \cellcolor[gray]{0.9} (3,6) \\
     \hhline{|=||=|=|=|}
     \cellcolor[gray]{0.9} (5,1) & 0 & $-\frac{11447}{104490}$ & $\frac{3491}{62694}$ \\
     \hline
     \cellcolor[gray]{0.9} (1,20) & $-\frac{11447}{104490}$ & 0 & $-\frac{3601}{1253880}$ \\
     \hline
    \cellcolor[gray]{0.9} (3,6) & $\frac{3491}{62694}$ & $-\frac{3601}{1253880}$ & 0 \\
    \hline
    \end{tabular}
    \hfill
    \begin{tabular}{|c||c|c|c|}
    \hline
      \cellcolor[gray]{0.9} $h_{i,j,6,3}$    & \cellcolor[gray]{0.9} (5,1) & \cellcolor[gray]{0.9} (1,20) &  \cellcolor[gray]{0.9} (3,6) \\
     \hhline{|=||=|=|=|}
      \cellcolor[gray]{0.9} (5,1) & 0 & $\frac{101}{129}$ & $\frac{58}{129}$ \\
     \hline
      \cellcolor[gray]{0.9} (1,20) &  $\frac{101}{129}$  & 0 &  $\frac{53}{86}$ \\
     \hline
     \cellcolor[gray]{0.9} (3,6) &  $\frac{58}{129}$  & $\frac{53}{86}$ & 0\\
    \hline
    \end{tabular}
    \hfill \null
\label{numericalcoeffdev63v2}
\end{table}
\begin{table}[H]
    \centering
    \begin{tabular}{|c||c|c|c|c|c|c|c|c|c|c|}
        \hline
         \cellcolor[gray]{0.9} &\cellcolor[gray]{0.9} $b_4$ &\cellcolor[gray]{0.9} $b_6$ &\cellcolor[gray]{0.9} $d_4$ &\cellcolor[gray]{0.9} $d_6$ &\cellcolor[gray]{0.9} $h_4$ &\cellcolor[gray]{0.9} $h_{6,1}$ &\cellcolor[gray]{0.9} $h_{6,2}$ &\cellcolor[gray]{0.9} $h_{7,1}$ &\cellcolor[gray]{0.9} $h_{7,2}$ &\cellcolor[gray]{0.9} $h_{7,3}$ \\ 
         \hhline{|=||=|=|=|=|=|=|=|=|=|=|}
         \cellcolor[gray]{0.9}(5,1)  & $-\frac{115}{129}$ & $\frac{9242120}{2476327}$  & $-\frac{34}{43}$ & $\frac{110905440}{2476327}$ & $-\frac{3}{2}$ & $\frac{5}{43}$ & $\frac{59}{16}$ & $-\frac{1}{5}$ & $-\frac{11}{160}$ & $-\frac{1}{10}$ \\
         \hline
         \cellcolor[gray]{0.9}(1,20)  & $-\frac{115}{129}$ & $\frac{2805280}{7428981}$ & $-\frac{34}{43}$ & $\frac{7872160}{2476327}$ & $\frac{3}{2}$ & $\frac{58}{129}$ & $\frac{39}{16}$ & $\frac{1}{10}$ & $-\frac{7}{160}$ & $-\frac{1}{10}$ \\
         \hline
         \cellcolor[gray]{0.9}(3,6) & $-\frac{115}{129}$ & $\frac{15265820}{7428981}$ & $-\frac{34}{43}$ & $\frac{14294640}{2476327}$ & 0 & $\frac{275}{516}$ & $\frac{3}{8}$ & $-\frac{1}{20}$ & $-\frac{9}{160}$ & $-\frac{1}{10}$\\
         \hline
    \end{tabular}
\label{numericalcoeffdev63v3}
\caption{Numerical coefficients obtained in the expansion of equations of motion in the \textbf{44} sector with a $SO(3)\times SO(6)$ breaking}
\end{table}
We used all the equations in (\ref{EOMperturb63}) to obtain the previous tables, except the second one as the system is overdetermined. It is an Einstein Equation, that was used as a constraint to verify the expansion.

We can now compute the renormalized action, taking into account the appropriate counterterms to renormalize the divergences occurring in the on-shell action (\ref{Sonshell63}) in the limit $\epsilon \rightarrow 0$. In total, we find the following counterterms
\begin{equation}\label{counterterms63}
    \begin{split}
        S_{\rm ct1} &= \int _{r=\epsilon} \rd t \sqrt{|h|}\,\Big( \frac{9}{10}\,\rho -\frac{3}{20}\,\rho^{5/9}+\frac{1}{20}\,\rho^{1/9}+\frac{1}{40}\,\rho^{-3/9} \Big) ~, \\
        S_{\rm ct2} &= \int _{r=\epsilon} \rd t \sqrt{|h|}\Big( -\frac{4}{405}\,(9 b_{i,4}+4 d_{i,4})-\frac{1}{1215}\,(36\, b_{i,4}+54\, b_{i,6}+25\,d_{i,4}+24\,d_{i,6}\Big) c_{x_i,2}(t,\epsilon)^2~.
    \end{split}
\end{equation}
Consequently, the renormalized action is given by
\begin{equation}\label{renormalized63}
        S_{\rm ren} = \underset{\epsilon \rightarrow 0}{\mathrm{lim}} \;(S_{\rm on-shell} + S_{\rm ct1} + S_{\rm ct2}) ~\propto ~ \int \rd t \,c_{x_i,2}(t)c_{x_i,5}(t) \,.
\end{equation}
Holographic two-point functions can now be computed as functional derivatives of this renormalized action. We will present the analysis in the next section.

\subsubsection{Holographic correlation functions} 

The renormalized action (\ref{renormalized63}) is the starting point for the computation of holographic two-point functions. 
Going to momentum space
\begin{equation}\label{Srenxi63}
        S_{\rm ren} \propto \int \rd t\, c_{x_i,2}(t)c_{x_i,5}(t)
        \propto \int \rd q \,\tilde{c}_{x_i,2}(q)\,\tilde{c}_{x_i,5}(q) 
        \,,
\end{equation}
the holographic two-point functions are obtained as functional derivatives of this action w.r.t.\ $\tilde{c}_{x_i,2}(q)$ or $\tilde{c}_{x_i,5}$ , depending on which of the coefficients is identified with the source coupling to the corresponding operator. For scalars in the ${\bf 44}$, this is the subleading coefficient $\tilde{c}_{x_i,5}$, corresponding to the associated operator dimension $\Delta_-$, determined by the dual supersymmetric matrix model, see the discussion in \cite{Ortiz:2014aja}.
The dependence of $\tilde{c}_{x_i,2}(q)$ on $\tilde{c}_{x_i,5}(q)$ (and vice versa) is determined by imposing regularity of the solution in the bulk. At linear order, this leads to a relation of the type
\begin{equation}\label{regularityrenxi63}
    \tilde{c}_{x_i,5}(q) = C_i(q)\,\tilde{c}_{x_i,2}(q) \,,
\end{equation}
with a proportionality factor $C_i(q)$ to be determined, which carries the full information about the two-point function.
Accordingly, the two-point function is expressed as
\begin{equation}\label{twopoinfunc63}
    \langle O_i(0) O_i(q) \rangle \propto C_i(q)^{-1} \,.
\end{equation}

In order to determine the coefficient $C_i(q)$, one has to determine the regular solutions for the linearized fluctuation equations (\ref{EOMperturb63}). After Fourier transformation in $t$, these become ordinary second order differential equations for the scalar fluctuations. The near horizon ($r=0$) expansion of the solution regular in the bulk ($r\rightarrow\infty$) yields the coefficient $C_i(q)$ via (\ref{regularityrenxi63}). In the absence of analytical solutions to the fluctuation equations, we will have to employ numerical methods. For this, we follow the approach taken in \cite{Ortiz:2014aja}, see also \cite{Berg:2001ty,Berg:2002hy}.

\begin{figure}[tb]
    \centering
    \includegraphics{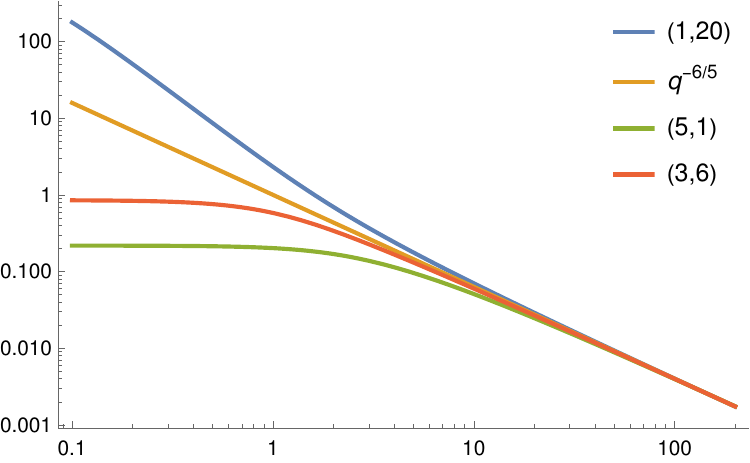}
    \caption{Numerical plot of the correlation functions $C_i(q)^{-1}$ for the operators dual to the ${\bf 44}$ scalar fields around the $SO(3)\times SO(6)$ deformation,
    together with the common asymptote $q^{-6/5}$ for $q\rightarrow \infty$, which is expected from the correlators (\ref{eq:corr1}) of the undeformed model.}
    \label{fig:plotratioxi631}
\end{figure}

After change of coordinates and fields 
\begin{equation}\label{fieldredefxi63}
    u = \sqrt{e^{3(r^{2/5})}-1} \,,\qquad\tilde{x}_i(u) \rightarrow u^2 \,\tilde{x}_i(u)
    \,,
\end{equation}
the fluctuation equations take the form
\begin{equation}\label{flucteqxi63}
    \begin{split}
        0&= \tilde{x}''_{(5,1)}(u) + \frac{2}{u}\frac{2u^2-1}{u^2+1}\,\tilde{x}'_{(5,1)}(u) - \frac{q^2u^3}{(u^2+1)^3}\,\tilde{x}_{(5,1)}(u)   \,,\\
        0&= \tilde{x}''_{(1,20)}(u) + \frac{2}{u}\frac{2u^2-1}{u^2+1}\,\tilde{x}'_{(1,20)}(u) - \Big( \frac{q^2u^3}{(u^2+1)^3}-2\frac{u^2-1}{(u^2+1)^2}\Big)\,\tilde{x}_{(1,20)}(u)  \,,\\
        0&= \tilde{x}''_{(3,6)}(u) + \frac{2}{u}\frac{2u^2-1}{u^2+1}\,\tilde{x}'_{(3,6)}(u) - \Big(\frac{q^2u^3}{(u^2+1)^3}-\frac{2u^2-1}{(u^2+1)^2}\Big)\,\tilde{x}_{(3,6)}(u)  \,,
    \end{split}
\end{equation}
where we have corrected an error in the last equation of equation (3.48) of \cite{Ortiz:2014aja}. In accordance with the expansion (\ref{eq:seriesXi}), (\ref{exprcoeffdev63relation}), the regular solution admits a boundary expansion as
\begin{equation}
        \tilde{x}_{i}(q,u) = 
        1 + \frac16\,(2\,h_{i,4}-3)\,u^2+\frac{\sqrt{3}}{9}\,C_i(q)\,u^3 + {\cal O}(u^4)\,,
\end{equation}
with the $q$-independent factor $h_{i,4}$ from  (\ref{exprcoeffdev63relation}). Upon further redefinition 
\begin{equation}\label{defYtildeXxi63}
    y_i(q,u) = \tilde{x}_i(q,u) + \frac{1}{3u}\frac{\rd \tilde{x}_i}{\rd u}(q,u)
    =1+\frac19\,(2\,h_{i,4}-3)\,\gamma_i+\frac{\sqrt{3}}{9}\,u\,C_i(q) + {\cal O}(u^2)
    \,,
\end{equation}
we arrive at a function which carries the sought-after coefficient $C_i(q)$ at linear order in $u$. Formulating the second-order differential equations (\ref{flucteqxi63}) in terms of $y_i$, we numerically determine two independent solutions $\mathring{y}_i$ and $\bar{y}_i$, uniquely fixed by the respective choice of initial conditions at $u=0$
\begin{equation}\label{yxi63solpart}
    \lbrace \mathring{y}_i(0) = 1 ~,~ \mathring{y}_i{}'(0) = 0 \rbrace ~,~ \lbrace \bar{y}_i(0) = 0 ~,~ \bar{y}_i{}'(0) = 1 \rbrace \,.
\end{equation}
For generic values of $q$, both solutions share the same divergent asymptotic behavior in the bulk  (i.e.\ for $u \rightarrow \infty$). The unique solution $y_i^{\rm (reg)}$ (\ref{defYtildeXxi63}), regular in the bulk, may be expressed as a linear combination of these two solutions
\begin{equation}\label{yregxi63}
    y_i^{\rm (reg)} = \left(1+\tfrac19\,(2\,h_{i,4}-3)\right)\mathring{y} + \frac{\sqrt{3}}{9}\,C_i(q) \,\bar{y}_i  \,.
\end{equation}
With both, $\mathring{y}_i$ and $\bar{y}_i$, dominating $y_i^{\rm (reg)}$ in the bulk, we may then read off the factor $C_i(q)$ in the limit
\begin{equation}\label{CILIMYxi63}
    C_i \propto 
    \underset{u \rightarrow \infty}{\textrm{lim}}\,\frac{\mathring{y}}{\bar{y}}
    \,,
\end{equation}
which can be computed numerically for each value of $q$. 
In Figure~\ref{fig:plotratioxi631}, we have plotted the (normalized)
correlation functions (\ref{twopoinfunc63}) for the three representations of scalar fields in log-log scales.

Their asymptotics at $q\rightarrow \infty$ is in good agreement with the scaling behavior $q^{-6/5}$, which is expected from the correlators (\ref{eq:corr1}) of the undeformed model. Numerically, we find the scaling behavior
\begin{equation}
\label{ABCcoeffxi63}
        C_{(5,1)}:\,q^{1.19}\,, \qquad
        C_{(1,20)} :\, q^{1.21}\,,\qquad 
        C_{(3,6)} :\, q^{1.20}\,.
\end{equation}

\begin{figure}[H]
    \centering
    \includegraphics{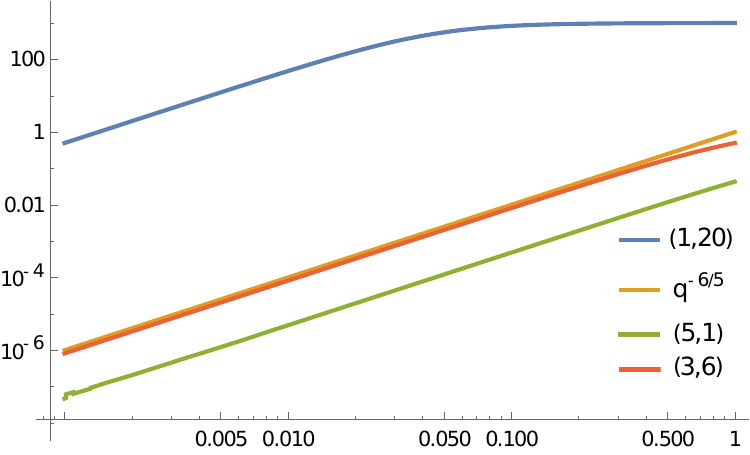}
    \caption{Numerical plot of $C_i^{-1}(q)-C_i^{-1}(0)$ for the operators dual to the scalar fields of the theory. This plot in log-log represents their values and comparison to $q^2$ in yellow.}
    \label{fig:plotratioxi63q0}
\end{figure}

Around $q=0$, we can analytically solve equations (\ref{flucteqxi63})
\begin{equation}\label{solutionq036}
    \begin{split}
        \tilde{x}_{(5,1)} &= \frac{1}{8}\Big(\frac{u(u^2-1)}{(u^2+1)^2} + \arctan(u) \Big)c_1 + c_2~\\
        \tilde{x}_{(1,20)} &= \frac{c_1}{u^2+1}+c_2\frac{u-\arctan(u)}{u^2+1}~\\
        \tilde{x}_{(3,6)} &=  \frac{c_1}{\sqrt{1+u^2}}+c_2\frac{u^2 \arctan(u) + \arctan(u)-u}{2(1+u^2)^{3/2}}~,
    \end{split}
\end{equation}
with integration constants $c_1$ and $c_2$. 
Expanding these solutions in series around $u = 0$, we find $C_{(1,20)}|_{q=0}$ vanishing, whereas $C_{(5,1)}|_{q=0}$ and $C_{(3,6)}|_{q=0}$ converge to  positive constants. For all functions, we find the universal asymptotic behavior
\begin{equation}
    C_i^{-1}(q)-C_i^{-1}(0) \propto q^2\,,
    \label{eq:q0q2}
\end{equation}
around $q=0$, as we illustrate in Figure~\ref{fig:plotratioxi63q0}. We will encounter the same asymptotic behavior in all correlators presented in the rest of the paper.

\subsection{Holographic correlation functions for the active scalar}

In this subsection we extend the analysis to the fluctuations of the active scalar $x$, in order to compute correlation functions of the associated singlet operator. In the asymptotic expansion
(\ref{frhoback63v2}), these fluctuations are captured by additional free coefficients $c_{x,0}(t)+c_{x,3}(t)\,r^{3/5}$ in the expansion of $x_{\rm fluc}$, left undetermined by the equations of motion in accordance with (\ref{exprcoeffdev63relation}).
At quadratic order, the fluctuations of $x$ cannot couple to the $x_i$, which we consequently set to zero. 
In contrast, the fluctuations of the active scalar couple to fluctuations in the dilaton-metric sector, i.e.\ also requires to consider sources in the functions $f$ and $\rho$, represented by the free coefficient $c_{f,7}$. Similar to the above analysis, we first compute the renormalized action, taking into account the counterterms to renormalize the divergences occurring in the on-shell action (\ref{Sonshell63}) in the limit $\epsilon \rightarrow 0$. With the following counterterms
\begin{equation}\label{counterterms63xx}
    \begin{split}
    S_{\rm ct3} &= \int _{r=\epsilon} \rd t \sqrt{|h|}\Big( \frac{1}{12} \rho ^{-1/3} \left(64 c_{x,0}(t)^3-69 c_{x,0}(t)^2+5\right)+\frac{3}{5} \rho^{1/9} \left(c_{x,0}(t)^2-1\right) \Big )\,,
    \end{split}
\end{equation}
we obtain the renormalized action given by
\begin{equation}\label{renormalized63xx}
    \begin{split}
        S_{\rm ren} &= \underset{\epsilon \rightarrow 0}{\mathrm{lim}} (S_{\rm on-shell} + S_{\rm ct1} + S_{\rm ct3}) ~,\\
        &\propto \int \rd t\, \Big ( \frac{7}{5760} c_{f,7}(t) + c_{x,0}(t) c_{x,3}(t) \Big) \,.
    \end{split}
\end{equation}
Again, we use the renormalized action in momentum space as a starting point for the holographic computation of the correponding two-point correlation functions
\begin{equation}\label{Srenxx63}
    \begin{split}
        S_{\rm ren} &\propto \int \rd q\, \Big( \tilde{c}_{x,0}(q)\tilde{c}_{x,3}(q) + \frac{7}{5760} \tilde{c}_{f,7}(q) \Big)~.
    \end{split}
\end{equation}
Similarly to the previous section, the holographic two-point functions are obtained as functional derivatives of this action w.r.t.\ $\tilde{c}_{x,0}$ or $\tilde{c}_{x,3}$, depending on which of the coefficients is identified with the source coupling to the corresponding operator. As the active scalar $x$ is again in the ${\bf 44}$, we obtain an analogous expression for the two-point correlation function as
\begin{equation}\label{twopoinfuncxx63}
    \langle O_x(0) O_x(q) \rangle \propto C_x(q)^{-1} \,,
\end{equation}
where $C_x(q)$  is the coefficient of proportionality
\begin{equation}\label{regularityrenxx63}
    \tilde{c}_{x,3}(q) = C_x(q)\,\tilde{c}_{x,0}(q) \,,
\end{equation}
for the function regular in the bulk. We can also try to compute an analogous coefficient $C_f(q)$ that is not a correlator of a scalar of the \textbf{44}, but that we can try to compute similarly:
\begin{equation}\label{regularityrenxf63}
    \tilde{c}_{f,7}(q) = C_f(q)\,\tilde{c}_{x,0}(q) \,,
\end{equation}
The first step for the calculation of $C_x(q)$  is to solve a system of equations of motion (\ref{EOMperturb63}) linearized around the background (\ref{dsweyl63}).
After taking the Fourier transform with respect to time, we are left with a system of three second order differential equation in the radial coordinate $r$, for the fluctuations $\delta x$, $\delta f$ and $\delta \rho$. There exists a unique solution that is regular in the bulk (i.e.\ falls off sufficiently fast as $r$ goes to infinity). The $C_x(q)$ can be read off from the 
power series expansion of this regular solution via (\ref{regularityrenxx63}). For computational convenience, we will make employ the same change of coordinates as in (\ref{fieldredefxi63})
\begin{equation}\label{fieldredefxx63}
    u = \sqrt{e^{3(r^{2/5})}-1} \,.
\end{equation}
The linearized fluctuation equations then take the form
\begin{equation}\label{flucteqxx63}
    \begin{split}
        0= &~\delta\rho''(u) -\frac{3 \left(\left(u^2+3\right) \delta f'(u)+\left(u^2+7\right) \delta\rho'(u)\right)}{2 \left(u^3+u\right)}-\frac{q^2 u^3 \delta\rho (u)}{\left(u^2+1\right)^3}\\
    &-\delta x(u)\frac{2 \left(u^2+5\right)  \log \left(u^2+1\right)}{\left(u^2+1\right)^2} ~,\\
        0= &~\delta f''(u)+\frac{2 \left(u^2-6\right) \delta f'(u)+2 \left(u^2+3\right) \delta\rho '(u)+3 u^2 \log \left(u^2+1\right) \delta x'(u)}{3 \left(u^3+u\right)}\\
        &-\delta x(u)\frac{ (\left(u^2+5\right) \log \left(u^2+1\right)-18 u^2)}{9 \left(u^2+1\right)^2}\,,\\
        0= &~\delta x''(u)+\frac{2 \left(u^2 \delta f'(u)+u^2 \delta\rho'(u)+\left(2 u^2-3 \log \left(u^2+1\right)\right) \delta x'(u)\right)}{u \left(u^2+1\right) \log \left(u^2+1\right)}\\
        &+\delta x(u) \frac{\left(\left(-3 q^2 u^5+8 u^6+30 u^4+52 u^2+30\right) \log \left(u^2+1\right)-6 u^2 \left(u^4+6 u^2+5\right)\right)}{3 u^2 \left(u^2+1\right)^3 \log \left(u^2+1\right)},
    \end{split}
\end{equation}
as well as a relation
\begin{equation}\label{chg2xx63}
    \delta  \rho '(u) = -\frac{u\,\textrm{Log}[1+u^2]}{1+u^2} \, \delta x (u) \,.
\end{equation}
The form of the equations suggests the change of variables 
\begin{equation}\label{chg1xx63}
    \delta x \rightarrow \frac{1}{\textrm{Log}[1+u^2]}\delta y \,,
\end{equation}
in terms of which equations (\ref{flucteqxx63}), (\ref{chg2xx63}) reduce to
\begin{equation}\label{flucteqxx63v3}
    \begin{split}
        0&=\delta f''(u)+\frac{2 \left(u^2-6\right) \delta f'(u)+3 u^2 \delta y'(u)}{3 u \left(u^2+1\right)}-\frac{\left(7 u^2+23\right) \delta y(u)}{9 \left(u^2+1\right)^2}   ~,\\
        0&= \delta y''(u)+\frac{2 u^2 \delta f'(u)-6 \delta y'(u)}{u \left(u^2+1\right)}+\frac{\delta y(u) \left(-3 q^2 u^5+6 u^6+28 u^4+52 u^2+30\right)}{3 u^2 \left(u^2+1\right)^3} ~. 
        \end{split}
\end{equation}
Noting that $\delta f $ only appears under derivative, we introduce a new function
\begin{equation} 
H(u) = \delta f'(u) + \frac{u}{1+u^2}\delta y(u)\,,
\end{equation}
together with a field redefinition $\delta y\rightarrow u^2\delta y$. The system then becomes
\begin{equation}\label{flucteqxx63v5}
    \begin{split}
         0&=H'(u) +\frac{2(u^2-6)}{3u(u^2+1)}H(u)-\frac{12u^2(u^2-1)}{9(u^2+1)^2} \delta y(u)  ~,\\
        0&=u^2\delta y''(u)+ \frac{u(4 u^2-2)}{u^2+1}\delta y'(u)+2\frac{u}{u^2+1} H(u) +\frac{u^2 \left(-3 q^2 u^3+6 u^4+4 u^2-2\right)}{3(u^2+1)^3} \delta y(u) ~. 
        \end{split}
\end{equation}
Using the first equation to eliminate $\delta y$, we finally arrive at a third order differential equation for $H(u)$, given by
\begin{equation}\label{flucteqxx63v7}
\begin{split}
     0= &~\frac{9u^3(u^2+1)^2}{4(u^2-1)}H^{(3)}(u)+\frac{3}{4}\frac{(41 u^4-68 u^2+3)(u^2+1)u^2}{(u^2-1)^2}H''(u)\\
     &+3u^3\frac{-3 q^2 u^7+6 q^2 u^5-3 q^2 u^3+148 u^8-300 u^6-60 u^4+396 u^2+8}{4u(u^2-1)^3(u^2+1)} H'(u)  \\
     &+u^2\frac{-33 q^2 u^9+75 q^2 u^7-51 q^2 u^5+9 q^2 u^3+404 u^{10}-848 u^8+16 u^6+1640 u^4+348 u^2-24}{4(u^2-1)^3(u^2+1)^2}H(u)   \,,
\end{split}
\end{equation}
where we have done a final redefinition $H(u)\rightarrow u^3H(u)$. Upon expansion around $u=0$, the free coefficients of $H(u)$ are located in orders 0, 1 and 3, and in one-to-one correspondence with the free coefficients $c_{x,0}$, $c_{x,3}$ and $c_{f,7}$ in the fluctuation expansion. Similarly to the method presented in previous sections, it is useful to encode these coefficients in the function
\begin{equation}\label{chg4xx63}
    G(u) = H(u) + H'(u) ~.
\end{equation}
and its expansion
\begin{equation}
    G(u) = \alpha(q) + \beta(q) u + \gamma(q) u^2 \,.
\end{equation}
The resulting equation of motion for $G$ can be solved numerically for given initial conditions at $u = 0$. Let $G_1$, $G_2$ and $G_3$ denote the unique solutions with initial conditions
\begin{equation}\label{yxi63solpartxx}
\begin{split}
        \lbrace G_0(0) &= 1 ~,~ G_0'(0) = 0,~ G_0''(0) = 0 \rbrace ~,\\~ \lbrace G_1(0) &= 0 ~,~ G_1'(0) = 1,~ G_1''(0) = 0 \rbrace ~,\\~ \lbrace G_2(0) &= 0 ~,~ G_2'(0) = 0,~ G_2''(0) = 1 \rbrace ~.
\end{split}
\end{equation}
respectively, then the unique solution $G_{\rm reg}$ regular in the bulk (when $u \rightarrow \infty$) may be written (up to a global normalization factor) as a linear combination
\begin{equation}\label{Gregxx63}
    G_{\rm reg} = G_0 + b_1(q) G_1 + b_2(q) G_2 ~.
\end{equation}
We can numerically determine $b_1(q)$ and $b_2(q)$ by computing limits of ratios of the numerical solution $G_0$, $G_1$ and $G_2$, analogous to (\ref{CILIMYxi63}) above. Putting everything together, we obtain
\begin{equation}
    C_x(q) = \frac{c_{x,3}(q)}{c_{x,0}(q)} \propto \frac{60+17b_1(q)+11b_2(q)}{b_1(q)-1} ~,~ C_f(q) = \frac{c_{f,7}(q)}{c_{x,0}(q)} \propto \frac{8+3b_1(q)}{b_1(q)-1} ~.
\end{equation}
which can be computed numerically for each value of $q$. 
From the renormalized action (\ref{Srenxx63}), we see that $C^{-1}_x(q)$ carries the holographic function, similar to what we have seen for the other scalars in the ${\bf 44}$. In contrast to those other scalars however, $C_f(q)$ carries the information about a non-vanishing one-point correlation function for the operator dual to this active scalar, just as expected for a vev deformation of the original model.

\begin{figure}[H]
    \centering
    \includegraphics{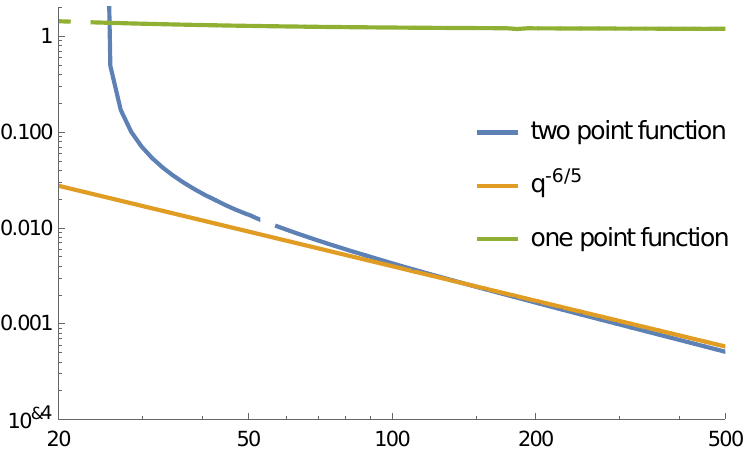}
    \caption{Numerical plot of the correlator for the active scalar. The two-point function $C^{-1}_x$ has an asymptote to $q^{-6/5}$. }
    \label{fig:plotratioxx631}
\end{figure}

In Figure~\ref{fig:plotratioxx631}, we have plotted the (normalized) correlation functions (\ref{twopoinfuncxx63}) in log-log scales. The asymptotics of $C^{-1}_x(q)$ carries the  at $q\rightarrow \infty$ is close to the scaling behavior $q^{-6/5}$, which is expected from the correlators (\ref{eq:corr1}) of the undeformed model. 
Numerically, we find the scaling behavior
\begin{equation}
\label{ABCcoeffxx}
        C_{x}:\,q^{1.26}\,, \qquad
        C_{f}:\,q^{0.03}\,.
\end{equation}
We note, that compared to the above analysis, the numerical accuracy is reduced,
due to the fact that we are dealing with third order fluctuation equations.

\subsection{On-shell action, renormalization, and correlators for scalars in the \textbf{84} sector}

In this subsection, we finally set up the holographic renormalization  and the computation of their two-point correlation functions for scalars in the \textbf{84} representation. The analysis is performed largely analogously to the discussion in section~\ref{subsec:2pt44} for the scalars in the ${\bf 44}$. The scalars in the ${\bf 84}$ decompose as given in (\ref{eq:4484to36}) under $SO(6)\times SO(3)$. However, the singlet scalar $(1,1)$ among the ${\bf 84}$, unlike its counterpart in the ${\bf 44}$ does not mix with the fluctuations of the dilaton-metric sector, so that in this case we can analyze all representations at once. 

\subsubsection{Asymptotic analysis}

The fluctuation equations in this sector are given by
\begin{equation}\label{eqmotionaxion63}
        0 = \rho^{-1/3}\nabla(\rho^{1/3}e^{-3x}\partial y_{i}) + \frac{1}{2}F_{y_{i}}(x) \,,
\end{equation}
with 
\begin{align}
F_{y_{(1,1)}}(x) =\,& 6e^{-x}\,,
\nonumber\\
F_{y_{(3,6)}}(x) =\,& -2e^{-4x}(1-4e^{3x})\,,
\nonumber\\
F_{y_{(3,15)}}(x) =\,& -2e^{2x}(-4+e^{3x})
\,,
\nonumber\\
F_{y_{(1,20)}}(x) =\,& 6e^{2x}
\,,
\end{align}
as found in (\ref{eq:LflucPhi}), (\ref{eq:KM88}). We parametrize the  fluctuations in this sector by the following expansion ansatz
\begin{equation}\label{axionfluc63}
    y_{i}(t,r) = c_{y_{i},1}(t)\,r^{1/5} + c_{y_{i},3}(t)\,r^{3/5} + c_{y_{i},5}(t)\,r + c_{y_{i},6}(t)\,r^{6/5} + c_{y_{i},7}(t)\,r^{7/5} + {\cal O}(r^{8/5})
    \,.
\end{equation}
Inserting this ansatz into the equations of motion (\ref{eqmotionaxion63}), we find that the independent coefficients in this expansion are $c_{y_i,1}$ and $ c_{y_i,3}$, while the remaining coefficients are determined as
\begin{equation}\label{exprcoeffyiexp63}
    \begin{split}
        c_{y_i,5} &= a_{i,1}\,c_{y_i,1} + a_{i,2}\,c_{y_i,3} \,,\\
        c_{y_i,6} &=  -\frac{1}{15}\,\Ddot{c}_{y_i,1} \,,\\
        c_{y_i,7} &= b_{i,1}\,c_{y_i,1} + b_{i,2}\,c_{y_i,3}  \,.
    \end{split}
\end{equation}
The numerical coefficients for the different representations are given by
\begin{table}[H]
    \centering
    \begin{tabular}{|c||c|c|c|c|}
        \hline
         \cellcolor[gray]{0.9} &\cellcolor[gray]{0.9} $y_{(1,1)}$ & \cellcolor[gray]{0.9} $y_{(3,6)}$ & \cellcolor[gray]{0.9} $y_{(3,15)}$& \cellcolor[gray]{0.9}$y_{(1,20)}$ \\ 
         \hhline{|=||=|=|=|=|}
         \cellcolor[gray]{0.9} $a_{i,1}$ & $\frac{9}{32}$ & $\frac{33}{32}$ & $-\frac{15}{32}$ & $-\frac{135}{32}$ \\
         \hline
         \cellcolor[gray]{0.9} $a_{i,2}$ & $\frac{24}{32}$ & $-\frac{3}{4}$ & $-\frac{9}{4}$ & $-\frac{30}{8}$ \\ 
         \hline
         \cellcolor[gray]{0.9} $b_{i,1}$ & $\frac{41}{192}$ & $-\frac{241}{192}$ & $\frac{125}{192}$ & $\frac{2435}{192}$ \\ 
         \hline
         \cellcolor[gray]{0.9} $b_{i,2}$ & $\frac{114}{192}$ & $\frac{27}{32}$ & $\frac{321}{96}$ & $\frac{777}{96}$ \\
         \hline
    \end{tabular}
\label{numericalcoeffaxion63}
\caption{Numerical coefficients obtained in the expansion of equations of motion in the \textbf{84} sector with a $SO(3)\times SO(6)$ breaking}
\end{table}

\subsubsection{Holographic correlation functions}

The holographic two-point functions are computed from the renormalized action, which is computed analogous to (\ref{Srenxi63}), and takes the form
\begin{equation}\label{Srenyi63}
        S_{\rm ren} \propto \int \rd t\, c_{y_i,1}(t)\,c_{y_i,3}(t) \propto \int \rd q\, \tilde{c}_{y_i,1}(q)\,\tilde{c}_{y_i,3}(q) \,,
\end{equation}
in momentum space.
In this sector, the holographic two-point functions are obtained as functional derivatives of this action w.r.t.\ the source $\tilde{c}_{y_i,1}(q)$, corresponding to the choice of $\Delta_+$ for the associated operator dimension, see the discussion in \cite{Ortiz:2014aja}.
The dependence of $\tilde{c}_{y_i,3}(q)$ on $\tilde{c}_{y_i,1}(q)$ is again determined by imposing regularity of the solution in the bulk. 
Accordingly, the two-point function of the dual operator is expressed as
\begin{equation}\label{twopoinfuncaxion63}
    \langle O_i(0) O_i(q) \rangle \propto C_{y_i}(q) \,,
\end{equation}
where $C_{y_i}(q)$ is the proportionality factor in
\begin{equation}\label{regularityrenaxion63}
    \tilde{c}_{y_i,3}(q) = C_{y_i}(q)\,\tilde{c}_{y_i,1}(q) ~.
\end{equation}
for the regular solution.

\begin{figure}[H]
    \centering
    \includegraphics{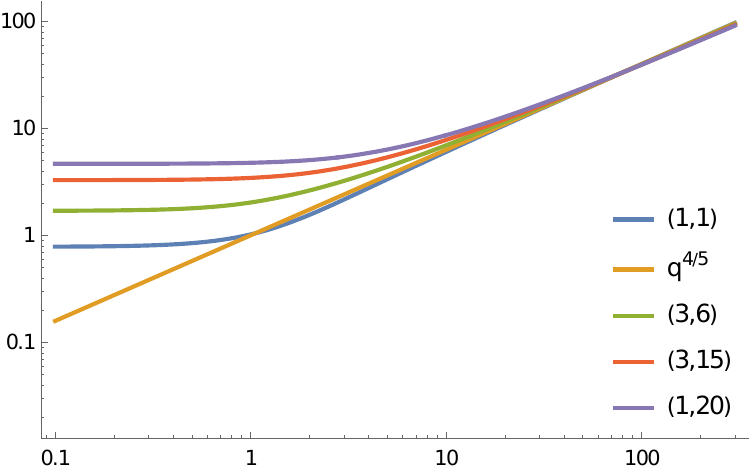}
    \caption{Numerical plot of the correlation functions $C_i(q)$ for the operators dual to the ${\bf 84}$ scalar fields around the $SO(3)\times SO(6)$ deformation. Their asymptotics at $q\rightarrow \infty$ approaches $q^{4/5}$, as expected from the correlators (\ref{eq:corr1}) of the undeformed model.}
    \label{fig:plotratioyi631}
\end{figure}

In the following, the function $C_{y_i}$ is determined numerically for each representation. We follow the same methods presented in  section~\ref{subsec:2pt44} above.
After change of coordinate and fields
\begin{equation}\label{fieldredefyi63v1}
    u = \sqrt{e^{3(r^{2/5})}-1} ~~,~~\tilde{y}_i(u) \rightarrow u~ \tilde{y}_i(u)
    \,,
\end{equation}
the fluctuation equations take the form
\begin{equation}\label{eqfluctaxion63}
    \begin{split}
        0 &= y_{(1,1)}''(u)+\frac{\left(u^2-1\right)}{u \left(u^2+1\right)} \,y_{(1,1)}'(u) +H_{{(1,1)}}\,y_{(1,1)}(u) \,,\\
        0 &= y_{(3,6)}''(u)+\frac{\left(3 u^2-1\right) }{u \left(u^2+1\right)}\,y_{(3,6)}'(u)+H_{{(3,6)}}\,y_{(3,6)}(u) \,,\\
        0 &= y_{(3,15)}''(u)+\frac{\left(5 u^2-1\right) }{u \left(u^2+1\right)}\,y_{(3,15)}'(u)+H_{{(3,15)}}\,y_{(3,15)}(u) \,,\\
        0 &=y_{(1,20)}''(u)+\frac{\left(7 u^2-1\right) }{u \left(u^2+1\right)}\,y_{(1,20)}'(u) +H_{{(1,20)}}\,y_{(1,20)}(u) \,,
    \end{split}
\end{equation}
for the different representations. The functions $H_{i}$ are given by
\begin{equation}\label{eqfluctaxion63H}
    \begin{split}
        H_{{(1,1)}} &=  \frac{ \left(-q^2 u^5-u^6+\left(6 (u^2+1)^{1/3}-7\right) u^2+3 \left((u^2+1)^{1/3}-1\right)+\left(3 (u^2+1)^{1/3}-5\right) u^4\right)}{u^2 \left(u^2+1\right)^3} \,,\\
        H_{{(3,6)}}  &= \frac{\left(-q^2 u^5+u^6-u^4+\left(4 (u^2+1)^{1/3}-5\right) u^2+3 \left((u^2+1)^{1/3}-1\right)\right)}{u^2 \left(u^2+1\right)^3} \,,\\
        H_{{(3,15)}}  &= \frac{ \left(-q^2 u^5+3 u^6+3 (u^2+1)^{1/3}-u^2 \left(3-2 (u^2+1)^{1/3}\right)-u^4 \left((u^2+1)^{1/3}-3\right)-3\right)}{u^2 \left(u^2+1\right)^3} \,,\\
        H_{{(1,20)}}  &= \frac{ \left(-q^2 u^5+5 u^6+7 u^4+3 (u^2+1)^{1/3}-u^2-3\right)}{u^2 \left(u^2+1\right)^3} \,.
    \end{split}
\end{equation}
All solutions admit an expansion
\begin{equation}
    \begin{split}
        \tilde{y}_{i}(q,u) = \alpha_i(q) +\beta_i(q)u^2 + {\cal O}(u^3)
        \,,
    \end{split}
\end{equation}
at $u=0$ (equivalent to $r=0$), and the ratio $C_{y_i}(q)$ is now given by
\begin{equation}\label{CIalphabetayi63}
   C_{y_i} \propto \frac{\beta_i(q)}{\alpha_i(q)} ~.
\end{equation}
The equations (\ref{eqfluctaxion63}), similarly to the fluctuation equations in the previous sections, can only be solved numerically. We implement the same procedure as above, in order to extract the ratio (\ref{CIalphabetayi63}). To begin with, let us introduce another function
\begin{equation}\label{defZtildeXyi63}
    z_i(q,u) = \tilde{y_i}(q,u) + \frac{1}{2}\frac{\rd \tilde{y_i}}{\rd u}(q,u)~,
\end{equation}
whose power expansion around $u=0$ goes as
\begin{equation}\label{devZyi63}
    z_i(q,u) = \alpha_i(q) + \beta_i(q)u + {\cal O}(u^2)
\end{equation}
For each representation, similarly to the previous sections we numerically solve the corresponding equation of motion for $z_i$ for given initial conditions at $u = 0$. Let $\mathring{z}_i$ and $\bar{z}_i$ be the unique solutions with the following initial conditions
\begin{equation}\label{Zyi63solpart}
    \lbrace \mathring{z}_i(0) = 1 ~,~\mathring{z}_i'(0) = 0 \rbrace ~,~ \lbrace\bar{z}_i(0) = 0 ~,~ \bar{z}_i'(0) = 1 \rbrace ~,
\end{equation}
Both are divergent in the bulk with the same asymptotic behavior, while the unique solution $z_i^{\rm (reg)}$ (\ref{defZtildeXyi63}), regular in the bulk ($u \rightarrow \infty$), may be expressed as a linear combination
\begin{equation}\label{zregyi63}
    z_i^{\rm (reg)} \propto \mathring{z}_i + \frac{\beta_i}{\alpha_i}(q) \,\bar{z}_i \,.
\end{equation}
With both, $\mathring{z}_i$ and $\bar{z}_i$, dominating $z_i^{\rm (reg)}$ in the bulk, we may then read off the factor $C_i(q)$ in the limit
\begin{equation}\label{CILIMYyi63}
    C_{y_i} \propto 
    \underset{u \rightarrow \infty}{\textrm{lim}}\,\frac{\mathring{z}}{\bar{z}}
    \,,
\end{equation}
which can again be computed numerically for each value of $q$. 
In Figure~\ref{fig:plotratioyi631}, we have plotted the (normalized)
correlation functions (\ref{twopoinfuncaxion63}) in log-log scales.

\begin{figure}[H]
    \centering
    \includegraphics{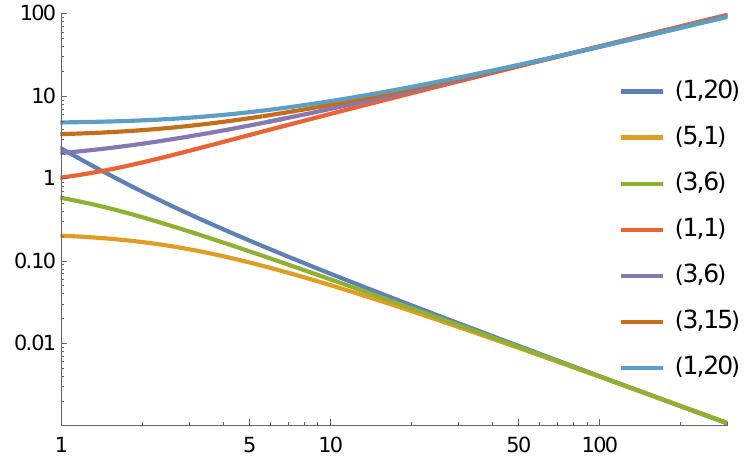}
    \caption{Plot of two-point correlators in both the \textbf{44} and the \textbf{84} sector. The $q\rightarrow\infty$ asymptotics reproduces the $q^{-6/5}$ and $q^{4/5}$ behavior of the undeformed model.}
    \label{fig:xiyi}
\end{figure}

Their asymptotics at $q\rightarrow \infty$ is in good agreement with the scaling behavior $q^{4/5}$, which is expected from the correlators (\ref{eq:corr1}) of the undeformed model. Numerically, we find the scaling behavior
\begin{equation}
\label{ABCcoeffyi63}
        C_{y_{(1,1)}}:\,q^{0.80}\,, \qquad
        C_{y_{(3,6)}} :\, q^{0.80}\,,\qquad 
        C_{y_{(3,15)}} :\, q^{0.80}\,,\qquad 
        C_{y_{(1,20)}} :\, q^{0.81}\,.
\end{equation}
For the asymptotics around $q=0$, we find the same behavior (\ref{eq:q0q2}) as in the \textbf{44} sector showed in Figure \ref{fig:plotratioxi63q0}. In Figure \ref{fig:xiyi}, we finally summarize all two-point correlators in the \textbf{44} and the \textbf{84} sector of the $SO(3)\times SO(6)$ deformed model.

\section{Generalisation to other \texorpdfstring{$SO(p)$}{SO(p)}~\texttimes \texorpdfstring{$~SO(9-p)$}{SO(9-p)} correlators}

We have in the previous section determined all holographic two-point correlation functions around the $SO(3)\times SO(6)$ domain wall solution. These should match the correlation functions in the vev deformed matrix model. In this section, we present the results for the holographic two-point functions for the other $SO(p)\times SO(9-p)$ domain wall solutions. Since the computation is done in full analogy to the previous case, we restrict to giving the final results as numerical plots. Also, we restrict to the scalar fields in the ${\bf 44}$ sector of the model, i.e.\ to correlation functions for the operators ${\cal O}_{44}$\,.

\subsection{Lagrangian for scalar fluctuations and general fluctuation equations}

We start by expanding the action (\ref{eq:gaugedLag}) to quadratic order in fluctuations around the $SO(p)\times SO(9-p)$ background (\ref{eq:solution1}), (\ref{eq:solution2}). After Weyl rescaling, and a change of coordinates and fields analogous to (\ref{fieldredefxi63})
\begin{equation}\label{fieldredefxipq}
    u = \sqrt{e^{\tfrac{9}{9-p}(r^{2/5})}-1} \,,\qquad\tilde{x}_i(u) \rightarrow u^2 \,\tilde{x}_i(u)
    \,,
\end{equation}
we find from (\ref{eq:LagXWeyl}) and (\ref{eq:KM44}) the following second order fluctuation equations
\begin{equation}\label{eqFlucpq2}
\begin{split}
     0 = &~ x_i''(u) + \frac{1}{u}\frac{(7-p) u^2-2}{ u^2+1}x'_i(u)\\
     &+\left(-\frac{q^2 u^3}{\left(u^2+1\right)^{(9-p)/2}}+\frac{F_i(u) \left(u^2+1\right)^{(2p-9)/9}-4 \left((p-4) u^2+5\right)}{2 \,u^2\left(u^2+1\right)}\right) x_i(u)  \,,
\end{split}
\end{equation}
where the $F_i(u)$ are obtained as $F_i=4M/\rho$ from (\ref{eq:KM44}) for the different representations.
These equations generalize the fluctuation equations  (\ref{flucteqxi63}) for $p=3$ to 
all other $SO(p)\times SO(9-p)$ domain wall solutions.
In the rest of this section, we spell out the fluctuation equations for the different values of $(p,9-p)$, and present the resulting numerical correlation functions, obtained by the same methods as in the previous section.

\subsection{Holographic correlation functions for the \texorpdfstring{$SO(p)$}{SO(p)}~\texttimes \texorpdfstring{$~SO(9-p)$}{SO(9-p)} solutions}

\subsubsection{The  \texorpdfstring{$SO(2)$}{SO(2)}~\texttimes \texorpdfstring{$~SO(7)$}{SO(7)} solution}
\begin{figure}[H]
        \centering
        \includegraphics{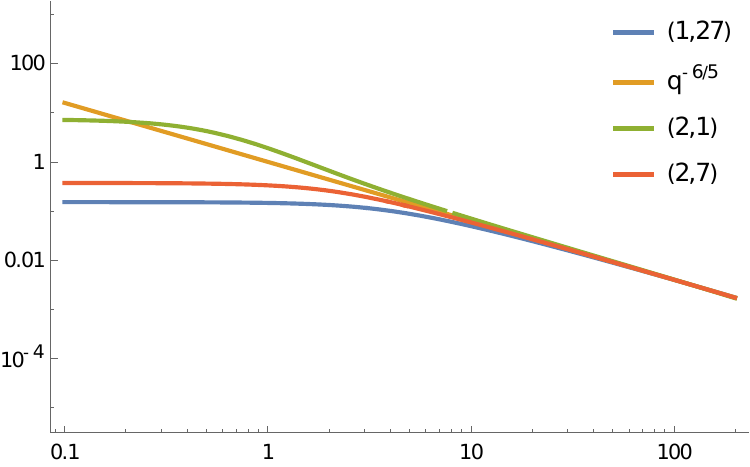}
        \caption{Numerical log-log plot of correlators of the $p=2$ case. The three functions have an asymptotic to $q^{-6/5}$ at $q\rightarrow \infty $ }
        \label{fig:2pt-p27}
\end{figure}

The scalars around the $SO(2) \times SO(7)$ domain wall break into the representations
\begin{equation}\label{44tofluc27}
\mathbf{44} \,\rightarrow (2,1) \oplus (2,7) \oplus (1,27)   \oplus (1,1)\,.
\end{equation}
The fluctuation equations read
\begin{equation}
    \begin{split}
        0&= \tilde{x}''_{(2,1)}(u) +\frac{5 u^2-2}{u^3+u}\tilde{x}'_{(2,1)}(u) + \left(-\frac{q^2 u^3}{\left(u^2+1\right)^{7/2}}+\frac{4 u^2-2}{\left(u^2+1\right)^2}\right) \tilde{x}_{(2,1)}(u)  ~,\\
        0&= \tilde{x}''_{(1,27)}(u) + \frac{5 u^2-2}{u^3+u}\tilde{x}'_{(1,27)}(u)-\frac{q^2 u^3}{\left(u^2+1\right)^{7/2}}\tilde{x}_{(1,27)}(u)  ~,\\
        0&= \tilde{x}''_{(2,7)}(u) +\frac{5 u^2-2}{u^3+u} \tilde{x}'_{(2,7)}(u)+ \left(-\frac{q^2 u^3}{\left(u^2+1\right)^{7/2}}+\frac{3 u^2-1}{\left(u^2+1\right)^2}\right) \tilde{x}_{(2,7)}(u) ~.
    \end{split}
\end{equation}
The resulting two-point correlators are presented in Figure~\ref{fig:2pt-p27}.

\subsubsection{The \texorpdfstring{$SO(4)$}{SO(4)}~\texttimes \texorpdfstring{$~SO(5)$}{SO(5)} solution}
\begin{figure}[H]
        \centering
        \includegraphics{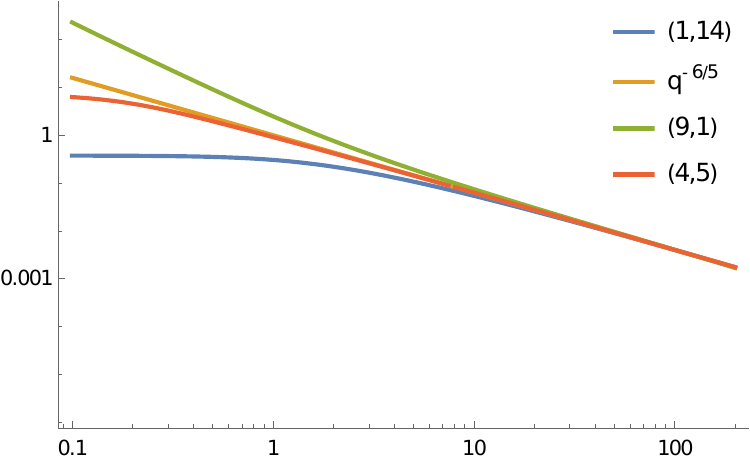}
        \caption{Numerical log-log plot of correlators of the $p=4$ case. The three functions have an asymptotic to $q^{-6/5}$ at $q\rightarrow \infty $ }
        \label{fig:2pt-p45}
\end{figure}

The scalars around the $SO(4) \times SO(5)$ domain wall break into the representations
\begin{equation}\label{44tofluc45}
\mathbf{44} \,\rightarrow (9,1) \oplus (4,5) \oplus (1,14)   \oplus (1,1)\,.
\end{equation}
The fluctuation equations read
\begin{equation}
    \begin{split}
        0&= \tilde{x}''_{(9,1)}(u) +\frac{3 u^2-2}{u^3+u}\tilde{x}'_{(9,1)}(u) + \left(-\frac{q^2 u^3}{\left(u^2+1\right)^{5/2}}-\frac{2}{\left(u^2+1\right)^2}\right) \tilde{x}_{(9,1)}(u)  ~,\\
        0&= \tilde{x}''_{(1,14)}(u) + \frac{3 u^2-2}{u^3+u}\tilde{x}'_{(1,14)}(u)-\frac{q^2 u^3}{\left(u^2+1\right)^{5/2}}\tilde{x}_{(1,14)}(u)  ~,\\
        0&= \tilde{x}''_{(4,5)}(u) +\frac{3 u^2-2}{u^3+u} \tilde{x}'_{(4,5)}(u)+ \left( -\frac{q^2 u^3}{\left(u^2+1\right)^{5/2}}+\frac{u^2-1}{\left(u^2+1\right)^2}\right) \tilde{x}_{(4,5)}(u) ~.
    \end{split}
\end{equation}
The resulting two-point correlators are presented in Figure~\ref{fig:2pt-p45}.

\subsubsection{The \texorpdfstring{$SO(5)$}{SO(5)}~\texttimes \texorpdfstring{$~SO(4)$}{SO(4)} solution} 
The scalars around the $SO(5) \times SO(4)$ domain wall break into the representations
\begin{equation}\label{44tofluc54}
\mathbf{44} \,\rightarrow (14,1) \oplus (5,4) \oplus (1,9)   \oplus (1,1)\,.
\end{equation}
The fluctuation equations read
\begin{equation}
    \begin{split}
        0= &~\tilde{x}''_{(1,9)}(u) +\frac{2 u^2-2}{u^3+u}\tilde{x}'_{(1,9)}(u)\\
        &+ \left(-\frac{q^2 u^3}{u^2+1}+\frac{-2 u^4+2 \left(5 \left(u^2+1\right)^{2/9}-6\right) u^2+10 \left(\left(u^2+1\right)^{2/9}-1\right)}{\left(u^3+u\right)^2} \right) \tilde{x}_{(1,9)}(u)  ~,\\
        0= &~\tilde{x}''_{(14,1)}(u) + \frac{2 u^2-2}{u^3+u}\tilde{x}'_{(14,1)}(u)\\
        &+\left(-\frac{q^2 u^3}{u^2+1}+\frac{-2 u^4+4 \left(2 \left(u^2+1\right)^{2/9}-3\right) u^2+10 \left(\left(u^2+1\right)^{2/9}-1\right)}{\left(u^3+u\right)^2}\right)\tilde{x}_{(14,1)}(u)  ~,\\
        0= &~\tilde{x}''_{(5,4)}(u) +\frac{2 u^2-2}{u^3+u} \tilde{x}'_{(5,4)}(u)\\
        &+ \left( -\frac{q^2 u^3}{u^2+1}+\frac{3 \left(3 \left(u^2+1\right)^{2/9}-4\right) u^2+10 \left(\left(u^2+1\right)^{2/9}-1\right)+\left(\left(u^2+1\right)^{2/9}-2\right) u^4}{\left(u^3+u\right)^2}\right) \tilde{x}_{(5,4)}(u) 
    \end{split}
\end{equation}

\begin{figure}[H]
        \centering
        \includegraphics{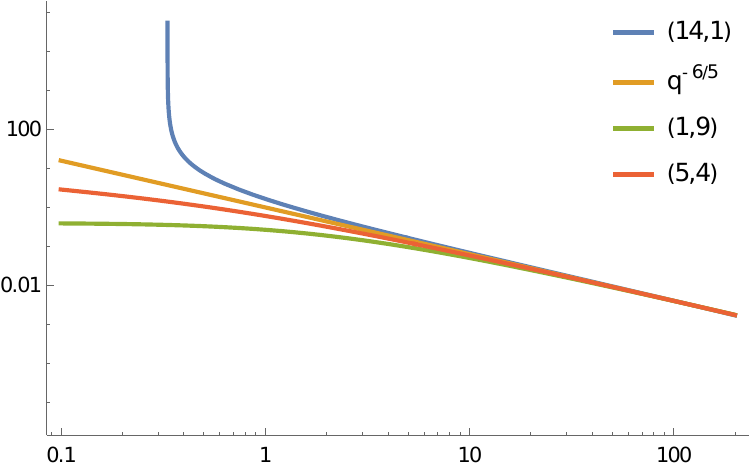}
        \caption{Numerical log-log plot of correlators of the $p=5$ case. The three functions have an asymptotic to $q^{-6/5}$ at $q\rightarrow \infty $ }
        \label{fig:2pt-p54a}
\end{figure}

\begin{figure}[H]
        \centering
        \includegraphics{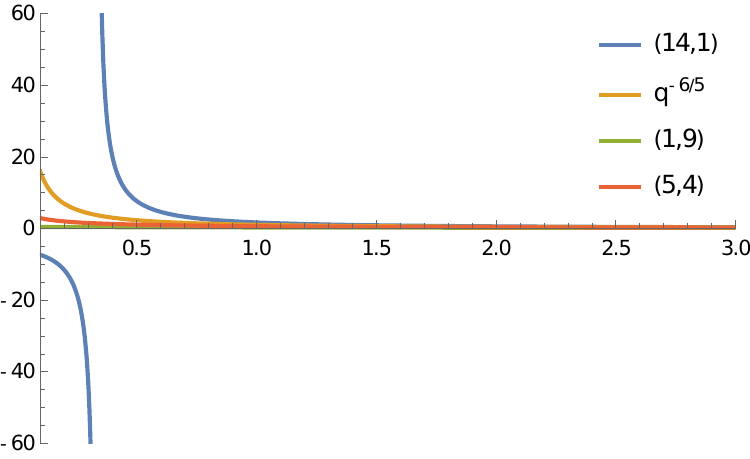}
    \caption{Numerical plot in normal unit of the $p=5$ case. The three functions have an asymptotic to $q^{-6/5}$ at $q\rightarrow \infty $. We see that the correlator (14,1) has a pole, and a negative value around $q=0$.  }
    \label{fig:2pt-p54b}
\end{figure}

The resulting two-point correlators are presented in log-log in Figure~\ref{fig:2pt-p54a}, and, in order to capture the intermediate pole, in ordinary units in Figure~\ref{fig:2pt-p54b}.


\subsubsection{The \texorpdfstring{$SO(6)$}{SO(6)}~\texttimes \texorpdfstring{$~SO(3)$}{SO(3)} solution} 
\begin{figure}[H]
        \centering
        \includegraphics{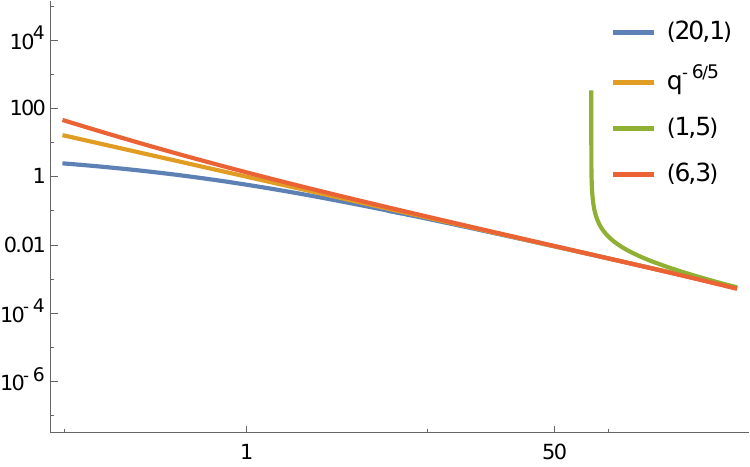}
        \caption{Numerical log-log plot of correlators of the $p=6$ case. The three functions have an asymptotic to $q^{-6/5}$ at $q\rightarrow \infty $ }
        \label{fig:2pt-p63a}
\end{figure}

\begin{figure}[H]
        \centering
        \includegraphics{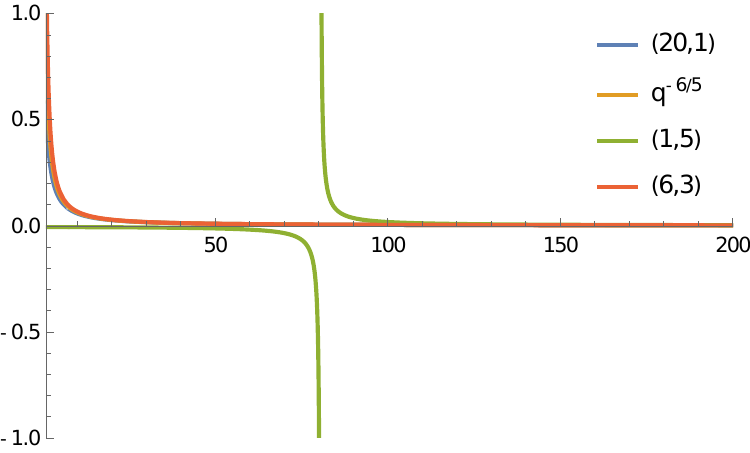}
        \caption{Numerical plot in normal unit of the $p=6$ case. The three functions have an asymptotic to $q^{-6/5}$ at $q\rightarrow \infty $. We see that the correlator (1,5) has a pole, and a negative value around $q=0$.  }
        \label{fig:plotratioxi633v2}
\end{figure}

The scalars around the $SO(6) \times SO(3)$ domain wall break into the representations
\begin{equation}\label{44tofluc63}
\mathbf{44} \,\rightarrow (20,1) \oplus (6,3) \oplus (1,5)   \oplus (1,1)\,.
\end{equation}
The fluctuation equations read
\begin{equation}
    \begin{split}
        0= &~\tilde{x}''_{(1,5)}(u)+\frac{u^2-2}{u^3+u}\tilde{x}'_{(1,5)}(u)+ \left(-\frac{q^2 u^3}{\left(u^2+1\right)^{3/2}}-\frac{4 u^2+26}{\left(u^2+1\right)^2}\right) \tilde{x}_{(1,5)}(u)  ~,\\
        0= &~\tilde{x}''_{(20,1)}(u) +\frac{u^2-2}{u^3+u}\tilde{x}'_{(20,1)}(u)-\frac{q^2 u^3}{\left(u^2+1\right)^{3/2}}\tilde{x}_{(20,1)}(u)   ~,\\
        0= &~\tilde{x}''_{(6,3)}(u)  +\frac{u^2-2}{u^3+u}\tilde{x}'_{(6,3)}(u) + \left(-\frac{q^2 u^3}{\left(u^2+1\right)^{3/2}}-\frac{1}{u^2+1}\right)\tilde{x}_{(6,3)}(u)  ~.
    \end{split}
\end{equation}
The resulting two-point correlators are presented in log-log Figure~\ref{fig:2pt-p63a}, and, in order to capture the intermediate pole, in ordinary units in Figure~\ref{fig:plotratioxi633v2}.


\subsubsection{The \texorpdfstring{$SO(7)$}{SO(7)}~\texttimes \texorpdfstring{$~SO(2)$}{SO(2)} solution} 
\begin{figure}[H]
        \centering
        \includegraphics{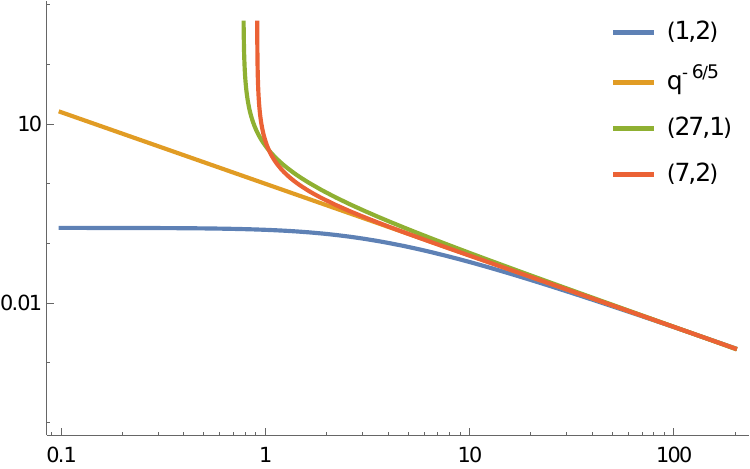}
        \caption{Numerical log-log plot of correlators of the $p=7$ case. The three functions have an asymptotic to $q^{-6/5}$ at $q\rightarrow \infty $ }

\label{fig:2pt-p72a}\end{figure}

\begin{figure}[H]
        \centering
        \includegraphics{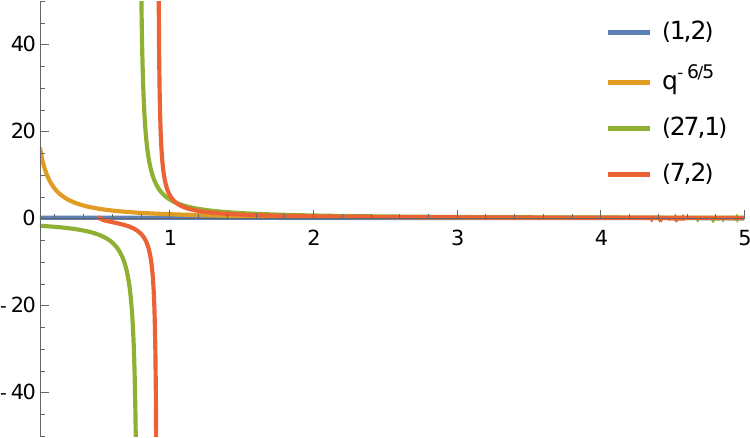}
        \caption{Numerical plot in normal unit of the $p=7$ case. The three functions have an asymptotic to $q^{-6/5}$ at $q\rightarrow \infty $. We see that the correlators (27,1) and (7,2) have a pole.  }
        \label{fig:2pt-p72b}
\end{figure}

The scalars around the $SO(7) \times SO(2)$ domain wall break into the representations
\begin{equation}\label{44tofluc72}
\mathbf{44} \,\rightarrow (27,1) \oplus (7,2) \oplus (1,2)   \oplus (1,1)\,.
\end{equation}
The fluctuation equations read
\begin{equation}
    \begin{split}
        0= &~\tilde{x}''_{(1,2)}(u) +\frac{-2}{u^3+u} \tilde{x}'_{(1,2)}(u) \\
        &+\left(-\frac{q^2 u^3}{\left(u^2+1\right)}+\frac{6 u^2 \left((u^2+1)^{1/9}-1\right)+10 \left((u^2+1)^{1/9}-1\right)-4 u^4 (u^2+1)^{1/9}}{u^4+u^2}\right) \tilde{x}_{(1,2)}(u)~,\\
        0= &~\tilde{x}''_{(27,1)}(u) +\frac{-2}{u^3+u}\tilde{x}'_{(27,1)}(u)\\
        &+\left(-\frac{q^2 u^3}{\left(u^2+1\right)}+\frac{\left(4 (u^2+1)^{1/9}-6\right) u^2+10 \left((u^2+1)^{1/9}-1\right)}{u^4+u^2}\right) \tilde{x}_{(27,1)}(u)  ~,\\
        0= &~\tilde{x}''_{(7,2)}(u) +\frac{-2}{u^3+u}\\ &+\left(-\frac{q^2 u^3}{\left(u^2+1\right)}+\frac{-10 \left((u^2+1)^{1/9}-1\right)+u^2 \left(6-5 (u^2+1)^{1/9}\right)+u^4 (u^2+1)^{1/9}}{u^4+u^2}\right)\tilde{x}''_{(7,2)}(u) ~.
    \end{split}
\end{equation}
The resulting two-point correlators are presented in log-log in Figure~\ref{fig:2pt-p72a}, and, in order to capture the intermediate poles, in ordinary units in \ref{fig:2pt-p72b}.


\section{Conclusion and perspectives}

In this paper, we study maximal supergravity in two dimensions, obtained from reduction of IIA supergravity on an $S^8$ sphere. We have constructed half-supersymmetric domain wall solutions preserving $SO(p)\times SO(9-p)$ subgroups of the original $SO(9)$ symmetry and determine their uplift to ten dimensions as well as the corresponding distributions of D0-branes. 
For the different domain walls, we have computed the scalar fluctuation equations and numerically extracted the holographic two-point correlation functions in the Coulomb branch of the matrix model for all scalars transforming in the ${\bf 44}$ of $SO(9)$.
In some of the cases, specifically for $p=5,6,7$, some of the resulting two-point functions exhibit poles. This is reminiscent of the observations in \cite{Freedman:1999gk} for D3-branes and merits further investigation. In particular, these results should be compared to independent computations using Monte Carlo methods directly on the matrix quantum mechanics side. In principle, our methods extend to the computation higher point correlators which could be explored similar to recent results in the undeformed BFSS model \cite{Biggs:2025xxx}.

It would also be interesting to further explore the difference between the domain wall solutions we have labeled  $SO(p) \times SO(q)$ as compared to $SO(q) \times SO(p)$. In the bulk, we have seen that these are different analytical continuations of the same gravitational solution, cf.\ (\ref{eq:flipPQ}), yet they give rise to a very different behavior of the associated holographic correlators.~A better understanding of their different boundary interpretations would be desirable.
Another interesting generalizaton to explore is the construction of similar solutions in the presence of non-vanishing axions, such as the spherical brane solutions of  
\cite{Bobev:2018ugk,Bobev:2024gqg}.

Two-dimensional gravity is of particular interest because, unlike in four spacetime dimensions, the Euclidean path integral can be well-defined in this case. Moreover, dilaton-gravity models such as JT gravity and its supersymmetric generalizations -- see, e.g. \cite{Mertens:2022irh} for a review -- provide solvable models of (super)gravity that capture the dynamics near the horizon of near-extremal black holes. It is plausible that at least certain universal features of such models could be obtained from specific truncations of 2d gauged supergravities, such as those considered in the present paper.~Establishing such a connection would be very interesting.

\subsection*{Acknowledgements}

We wish to thank
Franz Ciceri, Gianluca Inverso, and Ioannis Papadimitriou for interesting discussions and comments. A.B. thanks the Padova division of the Nuclear Physics institute of the INFN, and the University of Padova, in particular Gianguido Dall'Agata and Gianluca Inverso for hospitality while part of this work was done.

\appendix

\section{Matrix model and the holographic map}\label{sec:appendix}

In this appendix we review the holographic map in the context of the matrix model,  detailing the identification of our supergravity solutions with VEV deformations of the standard SO(9) vacuum of the BFSS at zero temperature.  

The BFSS matrix model \cite{deWit:1988ig,Banks:1996vh} consists of a set of adjoint U($N$) (or SU($N$), after decoupling the center of mass) bosonic and fermionic matrices $X_{I}$, $I=1, \dots, 9$, and $\psi_\alpha$, $\alpha=1, \dots, 16$, transforming in the fundamental and spinor representations of SO(9) respectively. 
Its action is given by,
\eq{\label{a1}
S=\frac{1}{g^2}\int\d t ~\!\text{Tr}\left(
\frac12\sum_I(D_tX_I)^2+\frac12\sum_\alpha\psi_\alpha D_t\psi_\alpha+\frac14\sum_{I,J}[X_I,X_J]^2
+\frac{i}{2}\sum_{I,\alpha,\beta}\psi_\alpha\Gamma^I_{\alpha\beta}[\psi_\beta,X_I]
\right)~,
}
where the SO(9) gamma matrices $\Gamma^I$ can be taken symmetric and real, 
and $D_t$ is a covariant derivative with respect to a (non-dynamical) gauge field in 0+1 dimensions. The ``Yang-Mills'' coupling constant $g$ is related to the  't Hooft coupling $\lambda$ and the  
string coupling $g_s$ via,
\eq{
\frac{\lambda}{N}=g^2=\frac{g_s}{4\pi^2 l_s^3}
~,}
where $l_s:=\sqrt{\alpha'}$ is the string length.

The holographic dual to the BFSS is the near horizon limit of $N$ coincident black D0 branes \cite{Itzhaki:1998dd}. In the string frame of IIA supergravity, the metric and dilaton are given by \cite{Horowitz:1991cd}, 
\eq{\spl{\label{a2}
\frac{\d s^2}{\alpha'}&=-f(r)~\!h(r)\d t^2+\frac{1}{f(r)}\left(\frac{\d r^2}{h(r)}+r^2\d\Omega_8^2\right)\\
h(r)&=1-\frac{r_h^7}{r^7}~;~~~f(r)=\left(\frac{r}{r_0}\right)^{\frac72}r_0^2\\
e^{-\phi}&=60~\!\pi^3N\left(\frac{r}{r_0}\right)^{\frac{21}{4}}~;~~~r_0=(240\pi^5\lambda)^{\frac13}
~,}}
where $\d\Omega_8^2$ is the metric of the unit eight-sphere $S^8$, and $r_h$ is  related to    the temperature of the black brane  via  $T\propto r_h^{5/2}$.~We will work in the  zero-temperature limit (extremal D0 branes), which amounts to taking $r_h\rightarrow 0$ and  setting  $h(r)=1$. 
The limits of validity of the 10d IIA description are set by the requirement that the effective string coupling is small ($e^\phi\ll1$) and  the effective curvature of $S^8$ is small in string units (which amounts to the condition $r\ll r_0$). 
Together, these two conditions give the allowed range of validity,\footnote{In the conventions of \eqref{a2}, $r$ has dimensions of inverse length. Redefining $r\rightarrow\alpha' r$, we can rewrite \eqref{a3} as \cite{Polchinski:1999br}, 
$$
 N^{\frac17}\ll \frac{r}{l_P} \ll N^{\frac13}~,
$$
where $l_P=g_s^{\frac13}l_s$ is the Planck length. 
}
\eq{\label{a3}
g_s^{\frac13}N^{\frac17}\ll rl_s \ll g_s^{\frac13}N^{\frac13}
~,}
which, in particular, implies $N\gg1$. 
The   supergravity description breaks down for, 
\eq{\label{ab5}r\gtrsim r_0~.}
This issue can be  addressed by setting up the holographic map  with the boundary fields inserted at $r=r_0$, instead of the actual boundary of spacetime at $r\rightarrow\infty$ \cite{Biggs:2025xxx}. 
Let us define a new radial variable $z$ and a  Euclidean time $\tau$  by,
\eq{\label{az}
z=\frac25\left(\frac{r}{r_0}\right)^{-\frac52}~;~~~\tau=i~\!r_0~\!t
~.}
In the zero-temperature limit, the metric and dilaton \eqref{a2} take the form,
\eq{\label{a4}
 \d s^2= \left(\frac52~\! z\right)^{\frac35}
\d s^2_{\text{dual}}~;~~~e^{-\phi}=60~\!\pi^3N\left(\frac{5}{2}z\right)^{-\frac{21}{10}}
~,}
where we have introduced the {\it dual string frame} metric \cite{Boonstra:1998mp}, 
\eq{\label{a5}
\d s^2_{\text{dual}}= 
\alpha'\left[
\left(\frac25\right)^{\!2}~\!\frac{\d\tau^2+\d z^2}{z^2}+\d\Omega_8^2
\right]
~,}
in which the geometry is AdS$_2\times S^8$ in Poincar\'{e} coordinates.~However, this is not the  standard conformal holography setup, as we have performed a Weyl rescaling, and the dilaton is running. 
Nevertheless, there is a convenient mathematical trick \cite{Kanitscheider:2009as} that allows us to (formally) rephrase the non-conformal holographic map  as a standard holographic dictionary, 
by embedding the background \eqref{a4} into an AdS$_{1+d}\times S^8$ solution of  a $(9+d)$-dimensional supergravity, and  analytically continuing to,
 \eq{\label{a6}d=\frac{14}{5}
 ~,}
 see \cite{Biggs:2023sqw} for a recent discussion.  
 Specifically, scalar   fields $\varphi$ of mass $M$ propagating in the bulk of AdS are associated to boundary operators $\mathcal{O}$ of dimension $\Delta$, such that,
 \eq{\label{amass}
 M^2=\Delta(\Delta-d)
 ~,}
with $d$ given in \eqref{a6}. Moreover, the two-point correlator is given by,
 \eq{\label{correlators}
 \langle \mathcal{O}(t+\Delta t)\mathcal{O}(t)  \rangle\propto \frac{1}{|\Delta t|^{2\Delta+1-d}}
 ~.}
 The holographic map allows us to read off the dimensions   of the boundary operators $\mathcal{O}$ from  the asymptotic near-boundary behavior of 
 their bulk duals $\varphi$, see \cite{Skenderis:2002wp} for a review, 
 \eq{\label{aasym}
 \lim_{z\rightarrow z_b}\varphi(\tau,z)=z^{d-\Delta}\varphi_s(\tau)+\cdots+z^{\Delta}\varphi_v(\tau)+\cdots
 ~,}
 where $\varphi_s$ 
is the source for the boundary  operator dual to  $\varphi$ while $\varphi_v$  is its VEV.\footnote{\label{foot}We have assumed that $\Delta\geq d/2$, so that the first term in the series is the dominant one. Note however, that if the mass \eqref{amass} is in the  interval given by,
\eq{
\label{interval}
-\frac{d^2}{4}<M^2<-\frac{d^2}{4}+1
~,\nonumber}
it is possible to have $\Delta<d/2$, so  that  the places  of the   $\varphi_s$, $\varphi_v$ terms in \eqref{aasym} are interchanged \cite{Klebanov:1999tb}.} 
 %
 %
As discussed below \eqref{ab5}, $z_b$ is the radial coordinate of the  boundary of the supergravity region. It is  obtained from \eqref{az} for $r\simeq r_0$, which gives $z_b\sim\mathcal{O}(1)$.~On the other hand, in view of  \eqref{a3}, \eqref{az}, the time scales at which we probe the theory obey, 
\eq{\label{a12}
1\ll\Delta\tau\sim\lambda^{\frac13}\Delta t\ll N^{\frac{10}{21}}~.}
It follows that $z_b$ 
 is parametrically smaller than  $\Delta z\sim\Delta\tau$,  so that 
 we may take $z_b\rightarrow0$ without loss of generality \cite{Biggs:2025xxx}.

For the scalars arising from the  IIA supergravity  metric and three-form compactified on $S^8$, dual to the 
 matrix model operators, 
 \eq{\spl{\label{a16}
 T_l&\propto
  \text{Tr}\left(
X_{(I_1}\cdots X_{I_l)}
 \right)
 \\
 J_l&\propto
 \text{Tr}\left(
 [X_I,X_J]X_{(I_1}\cdots X_{I_l)}
 \right)
 ~,}}
 where it is understood that terms proportional to $\delta_{I_mI_n}$ are removed from the right-hand sides, 
 the dimensions  were given in \cite{Sekino:1999av,Hanada:2011fq},
 \eq{\label{a17}
 \Delta(T_l)=\frac{2}{5}~\!l,~\text{for}~l\geq2~;~~~ \Delta(J_l)=\frac{1}{5}~\!(2l+7),~\text{for}~l\geq1
 ~.}
In \cite{Ortiz:2014aja} we calculated the near-boundary behavior of the bulk scalars in the ${\bf 44}$, ${\bf 84}$ of SO(9), corresponding to 
the operators $T_{l=2}$, $J_{l=1}$ of \eqref{a16},\footnote{To compare with  the asymptotic expansions in   \cite[Eq.~(2.46)]{Ortiz:2014aja}  for  $y_n$ with $n=44,84$, note that 
$\varphi_{n}=e^{a_n\phi/2} y_n$ with $a_{44}=0$, $a_{84}=4/7$ \cite[Eq.~(2.11)]{Ortiz:2014aja} and $e^\phi=r^{21/10}$ \cite[Eq.~(2.10)]{Ortiz:2014aja}, where  the radial coordinate $r$ corresponds  to $z^2$ here.  It can also be verified directly that 
the equations of motion for $y_n$ in  \cite[Eq.~(2.12)]{Ortiz:2014aja} can be rewritten as Klein-Gordon equations for the fields $\varphi_{n}=e^{a_n\phi/2} y_n$ in AdS$_{1+d}$, with masses given by  
\eqref{a6}, \eqref{amass} and $\Delta_{44}=\frac45$, $\Delta_{84}=\frac95$.} 
 \eq{\spl{\label{aasym3}
 \lim_{z\rightarrow 0}\varphi_{44}(\tau,z)&=z^{\frac45}\varphi_{44,v}(\tau)+z^{2}\varphi_{44,s}(\tau)+\cdots\\
  \lim_{z\rightarrow 0}\varphi_{84}(\tau,z)&=z\varphi_{84,s}(\tau)+z^{\frac95}\varphi_{84,v}(\tau)+\cdots
 ~.}}
These are of the form  \eqref{aasym}, for $d$ given in \eqref{a6} and $\Delta_{44}=\frac45$, $\Delta_{84}=\frac95$,  in agreement with the general formula \eqref{a17}. 
Note that $\Delta_{44}<d/2$ corresponds to the case discussed in Footnote \ref{foot}.

For the deformed backgrounds that we consider in this paper, which break the SO(9) isometry  to SO$(p)\times$ SO$(q)$, the external two-dimensional  part of the geometry asymptotes that of \eqref{a5},  
\eq{\label{a35}
 \rd \hat{s}_2^2 \; \underset{z \rightarrow 0}{\sim}\;
\frac{\d\tau^2+\d z^2}{z^2}     \left(1+{\cal O}(z^{\frac45})\right)
~,}
where we rewrote 
\eqref{AdS263} in  Poincar\'{e} coordinates by setting $r=z^2$, and Wick-rotated to Euclidean signature. 
The deformation of the background metric thus scales as $z^{\frac45}$ near the boundary. 
Comparing with \eqref{aasym3}, we therefore interpret it as a VEV deformation.

\newcommand\eprintarXiv[1]{\href{http://arXiv.org/abs/#1}{arXiv:#1}}

\providecommand{\href}[2]{#2}\begingroup\raggedright\endgroup


\end{document}